\documentclass[12pt]{article}
\RequirePackage{amsthm,amsmath,amsbsy,amsfonts}
\usepackage{bm,graphicx,psfrag,epsf}
\usepackage{enumerate}
\usepackage{natbib}
\usepackage{paralist, booktabs, multirow}
\usepackage{placeins} 
\usepackage{array}
\usepackage{makecell}

\newcommand{\blind}{0}

\numberwithin{equation}{section}
\theoremstyle{plain}

\usepackage{amssymb}
\usepackage[usenames]{color}
\usepackage[affil-it]{authblk}

\usepackage[mathscr]{eucal}

\usepackage{stmaryrd}

\usepackage{xspace}

\usepackage{paralist}

\usepackage{ifpdf}

\usepackage{graphicx}
\ifpdf
\usepackage{epstopdf}
\DeclareGraphicsExtensions{.eps,.pdf}
\else
\usepackage{epsfig}
\fi

\usepackage{booktabs}

\ifpdf
\usepackage[colorlinks,urlcolor=blue,citecolor=blue,linkcolor=blue]{hyperref}
\else
\newcommand{\href}[2]{{#2}}
\usepackage{url}
\fi

\usepackage{tabulary}

\usepackage{algorithm}
\usepackage{algpseudocode}

\ifpdf
\newcommand{\Sec}[1]{\hyperref[sec:#1]{Section~\ref*{sec:#1}}} 
\newcommand{\App}[1]{\hyperref[sec:#1]{Appendix~\ref*{sec:#1}}} 
\newcommand{\Eqn}[1]{\hyperref[eq:#1]{{\rm (\ref*{eq:#1})}}} 
\newcommand{\Part}[1]{\hyperref[part:#1]{(\ref*{part:#1})}} 
\newcommand{\Fig}[1]{\hyperref[fig:#1]{Figure~\ref*{fig:#1}}} 
\newcommand{\Tab}[1]{\hyperref[tab:#1]{Table~\ref*{tab:#1}}} 
\newcommand{\Thm}[1]{\hyperref[thm:#1]{Theorem~\ref*{thm:#1}}} 
\newcommand{\Lem}[1]{\hyperref[lem:#1]{Lemma~\ref*{lem:#1}}} 
\newcommand{\Prop}[1]{\hyperref[prop:#1]{Proposition~\ref*{prop:#1}}} 
\newcommand{\Cor}[1]{\hyperref[cor:#1]{Corollary~\ref*{cor:#1}}} 
\newcommand{\Def}[1]{\hyperref[def:#1]{Definition~\ref*{def:#1}}} 
\newcommand{\Alg}[1]{\hyperref[alg:#1]{Algorithm~\ref*{alg:#1}}} 
\newcommand{\Ex}[1]{\hyperref[ex:#1]{Example~\ref*{ex:#1}}} 
\newcommand{\As}[1]{\hyperref[as:#1]{Assumption~{\rm\ref*{as:#1}}}} 
\newcommand{\Reg}[1]{\hyperref[as:#1]{Condition~\ref*{reg:#1}}} 
\newcommand{\AlgLine}[2]{\hyperref[alg:#1]{line~\ref*{line:#2} of Algorithm~\ref*{alg:#1}}}
\newcommand{\AlgLines}[3]{\hyperref[alg:#1]{lines~\ref*{line:#2}--\ref*{line:#3} of Algorithm~\ref*{alg:#1}}}
\else
\newcommand{\Sec}[1]{{Section~\ref{sec:#1}}} 
\newcommand{\App}[1]{{Appendix~\ref{sec:#1}}} 
\newcommand{\Eqn}[1]{{(\ref{eq:#1})}} 
\newcommand{\Part}[1]{{(\ref{part:#1})}} 
\newcommand{\Fig}[1]{{Figure~\ref{fig:#1}}} 
\newcommand{\Tab}[1]{{Table~\ref{tab:#1}}} 
\newcommand{\Thm}[1]{{Theorem~\ref{thm:#1}}} 
\newcommand{\Lem}[1]{{Lemma~\ref{lem:#1}}} 
\newcommand{\Prop}[1]{{Property~\ref{prop:#1}}} 
\newcommand{\Cor}[1]{{Corollary~\ref{cor:#1}}} 
\newcommand{\Def}[1]{{Definition~\ref{def:#1}}} 
\newcommand{\Alg}[1]{{Algorithm~\ref{alg:#1}}} 
\newcommand{\Ex}[1]{{Example~\ref{ex:#1}}} 
\newcommand{\As}[1]{{Assumption~\ref{as:#1}}} 
\newcommand{\Reg}[1]{{R~\ref{reg:#1}}} 
\newcommand{\AlgLine}[2]{{line~\ref{line:#2} of Algorithm~\ref{alg:#1}}}
\newcommand{\AlgLines}[3]{{lines~\ref{line:#2}--\ref{line:#3} of Algorithm~\ref{alg:#1}}}
\fi

\usepackage[normalem]{ulem}

\usepackage{comment}
\usepackage{xspace}
\definecolor{blue}{rgb}{0.2,0.5,0.7}
\definecolor{green}{rgb}{0.3,0.68,0.29}
\definecolor{purple}{rgb}{0.6,0.31,0.64}

\usepackage{mathtools}

\usepackage{algorithm}
\usepackage{algpseudocode}
\usepackage{tikz}
\usetikzlibrary{arrows}

\newcommand{\Real}{\mathbb{R}}

\newcommand{\Tra}{^{\sf T}} 
\newcommand{\Inv}{^{-1}} 

\newcommand{\amp}{\mathop{\:\:\,}\nolimits}

\newcommand{\prox}{\operatorname{prox}}

\newcommand{\E}{\mathcal{E}}

\algnewcommand\algorithmicinput{\textbf{INPUT:}}
\algnewcommand\INPUT{\item[\algorithmicinput]}
\algnewcommand\algorithmicoutput{\textbf{OUTPUT:}}
\algnewcommand\OUTPUT{\item[\algorithmicoutput]}

\begin{document}

\def\spacingset#1{\renewcommand{\baselinestretch}%
{#1}\small\normalsize} \spacingset{1}


\if0\blind
{
  \title{\bf The Why and How of Convex Clustering}
  \author[1]{Eric C. Chi\thanks{Email: echi@umn.edu (Corresponding author)}} 
  \affil[1]{School of Statistics, University of Minnesota}  
  \author[1]{Aaron J. Molstad}
  \author[2]{Zheming Gao}
  \author[3]{Jocelyn T. Chi}
  \affil[2]{Department of Industrial and Systems Engineering, University of North Carolina, Charlotte}
  \affil[3]{Division of Computational Health Sciences, University of Minnesota}
  \date{}
  \maketitle
} \fi

\if1\blind
{
  \bigskip
  \bigskip
  \bigskip
  \begin{center}
    {\LARGE\bf Title}
\end{center}
  \medskip
} \fi

\bigskip
\begin{abstract}
This survey reviews a clustering method based on solving a convex optimization problem. Despite the plethora of existing clustering methods, convex clustering has several uncommon features that distinguish it from prior art. The optimization problem is free of spurious local minima, and its unique global minimizer is stable with respect to all its inputs, including the data, a tuning parameter, and weight hyperparameters. Its single tuning parameter controls the number of clusters and can be chosen using standard techniques from penalized regression. We give intuition into the behavior and theory for convex clustering as well as practical guidance. We highlight important algorithms and discuss how their computational costs scale with the problem size. 
Finally, we highlight the breadth of its uses and flexibility to be combined and integrated with other inferential methods.
\end{abstract}

\noindent%
\vfill

\newpage
\spacingset{1.45} 


\section{INTRODUCTION}

Clustering is a fundamental problem in statistics, machine learning, and data science. Clustering aims to organize a collection of objects into groups so that members within the same group are similar and members in different groups are dissimilar.  Although the task is trivial to state, the volume of literature dedicated to solving it attests to its non-triviality \citep{Hartigan1975, KaufmanRousseeuw1990, Gordon1999, Xu2008, Mirkin2013, Jaeger2023}.

Clustering has typically been posed as a discrete optimization problem. Unfortunately, 
the majority of discrete optimization problems are inherently combinatorial and require searching through potential solution sets whose size grows exponentially fast in the number of observations. The classic $k$-means problem, for example, is NP-hard \citep{Aloise2009, Dasgupta2009}. 
Moreover, greedy and iterative algorithms for solving these problems can get trapped in suboptimal local minima. Additionally, the output can be highly sensitive to hyperparameter choices, initialization, or small perturbations in the data. Some of these challenges have been mitigated in part by strategies that include clever initializations \citep{Arthur2007} and annealing schemes that steer solutions away from local minima \citep{Zhou2010, XuLange2019, Chakraborty2020}. An alternative approach is to cast the clustering task as a convex optimization problem. We review one such convex formulation of clustering that addresses all of the above issues. Other convex models have been proposed as convex relaxations of $k$-means based on semidefinite programming \citep{Peng2007, Awasthi2015, Mixon2017}, but our focus is on the sum-of-norms (SON) form of convex clustering \citep{PelDeSuy2005, Lin2011, Hocking2011}.  
SON convex clustering can also be seen as a convex relaxation of $k$-means \citep{Lin2011} as well as single linkage hierarchical clustering \citep{Tan2015}. For the rest of this paper, convex clustering will refer to the SON formulation.

\begin{figure}[H] 
    \centering
    \includegraphics[width=\textwidth]{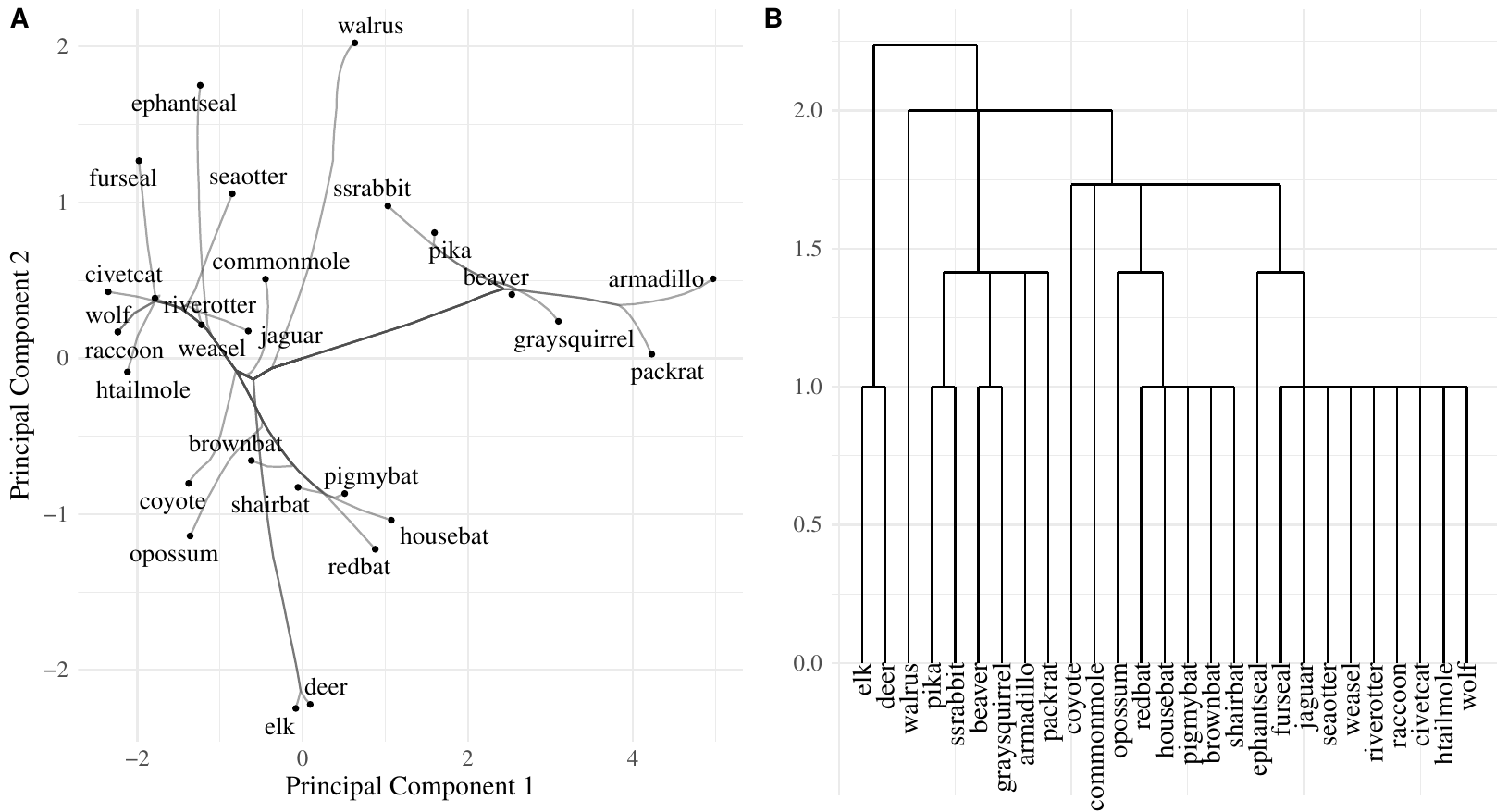}  
    \caption{Clustering trees obtained from (A) convex clustering and (B) hierarchical clustering with Euclidean distance single linkage.  Both methods produce similar clustering trees on the mammals dentition dataset.}
    \label{fig:trees_mammals}  
\end{figure}

To motivate why one might consider convex clustering in the face of so many clustering alternatives, we examine several illustrative examples.  We begin by comparing convex clustering to hierarchical clustering, a highly popular form of clustering in biological applications because of its ability to produce a dendrogram. We refer readers to \cite{Weylandt2019} for a comprehensive treatment and fast algorithm\footnote{Code is available at \url{https://github.com/DataSlingers/clustRviz}} for using convex clustering to visualize this popular tree organization. Figure \ref{fig:trees_mammals} shows the results of (A) convex clustering using \verb|cvxclustr| \citep{Chi2015} and (B) hierarchical clustering using \verb|hclust| in R on a mammals dataset featuring eight dentition variables \citep{de2009gifi}.  Like the dendrogram produced by hierarchical clustering in (B), convex clustering can also produce a similarly interpretable clustering tree as shown in (A).  We describe how one can obtain such a clustering tree later when we describe how a solution path emerges as a single smoothing parameter is varied.  In (A), the horizontal and vertical axes depict the first and second principal components of the mammals dataset.  In (B), the vertical axis shows the height at which clusters are merged.  For now, we simply point out that the clustering trees produced by both methods result in fairly similar clusterings on this mammals dataset, as visible from leaves corresponding to same branches.

\begin{figure}[H] 
    \centering
    \includegraphics[width=\textwidth]{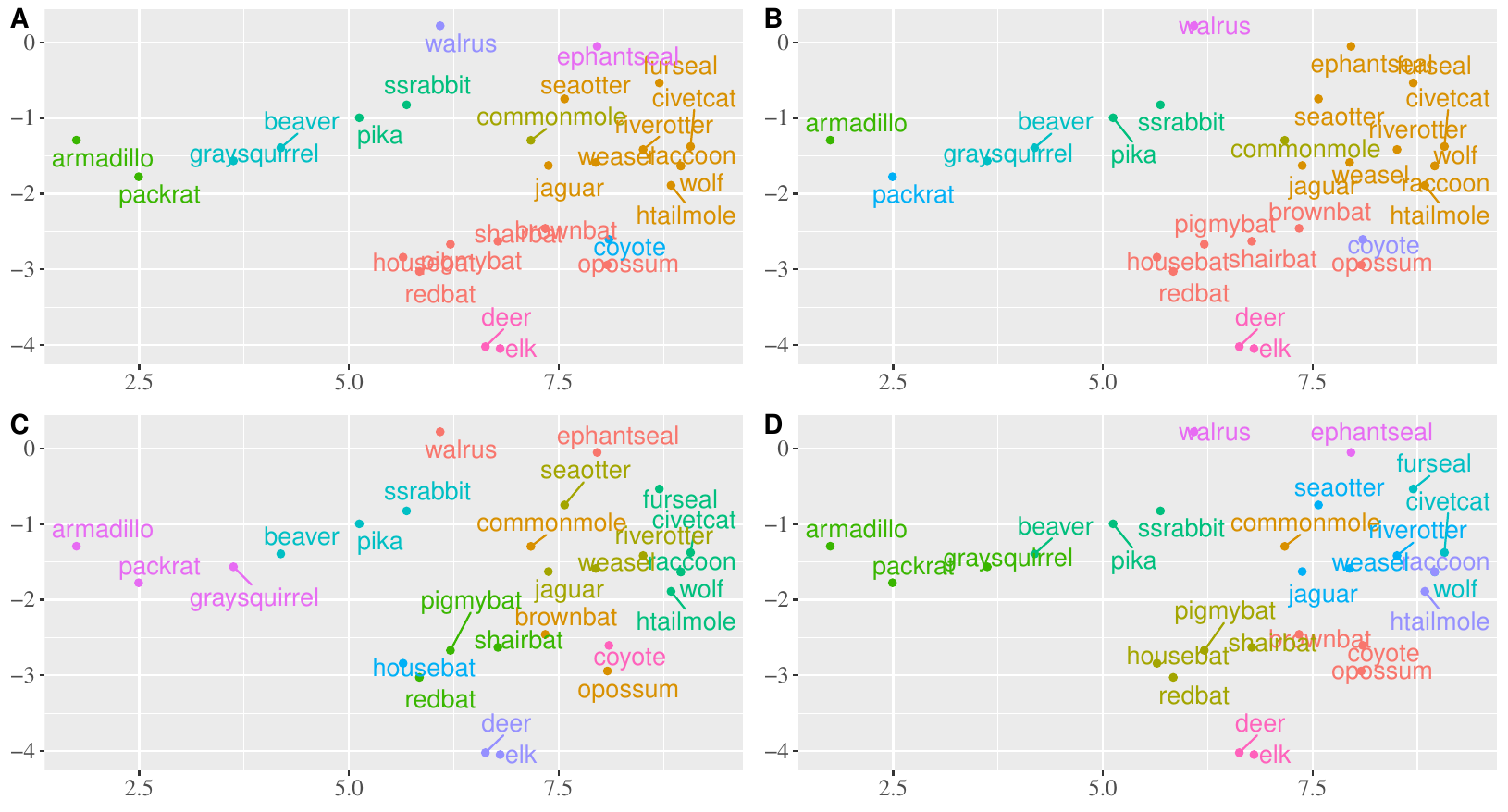}  
    \caption{Colors indicate the clustering memberships for the mammals dataset obtained from (A) convex clustering, (B) hierarchical clustering with Euclidean distance single linkage, (C) $k$-means clustering, and (D) Gaussian mixture models (GMM).  The number of clusters used in (B), (C), and (D) were matched to the number of clusters in (A).  On this dataset, convex clustering and hierarchical clustering produced similar clustering results.  By contrast, $k$-means produced results where not all bat variants were clustered together and GMM produced results where the coyote was clustered with the brown bat and the opossum.}
    \label{fig:comparisons_mammals}  
\end{figure}

To highlight differences in clustering results among popular statistical clustering methods, we additionally apply $k$-means clustering using \verb|kmeans| in R and Gaussian mixture models using \verb|mclust| \citep{mclust} on the same mammals dentition dataset.  Figure \ref{fig:comparisons_mammals} highlights how different clustering methods can produce very different clustering results.  The subplots depict clustering results on the same mammals dentition dataset obtained using (A) convex clustering, (B) hierarchical clustering, (C) $k$-means clustering, and (D) Gaussian mixture models (GMM).  The methods in (B), (C), and (D) were obtained using the same number of clusters as in (A).  The horizontal and vertical axes depict the first and second principal components of the mammals dataset.  Colors indicate clustering membership in each subplot and variations in coloring between subplots is due to variations in software labeling.  

While convex and hierarchical clustering produced similar results on the mammals dentition dataset, $k$-means and GMMs produced substantially different results.  For example, convex and hierarchical clustering clustered all bat types into a single group.  Meanwhile, $k$-means clustering grouped the same bats into three subgroups and GMM subdivided them into two groups.  As another example, convex, hierarchical, and $k$-means clustering similarly cast the coyote in its own cluster.  By contrast, GMM clustered the coyote into the same group as the brown bat and the opossum.

\begin{figure}[H] 
    \centering
    \includegraphics[width=\textwidth]{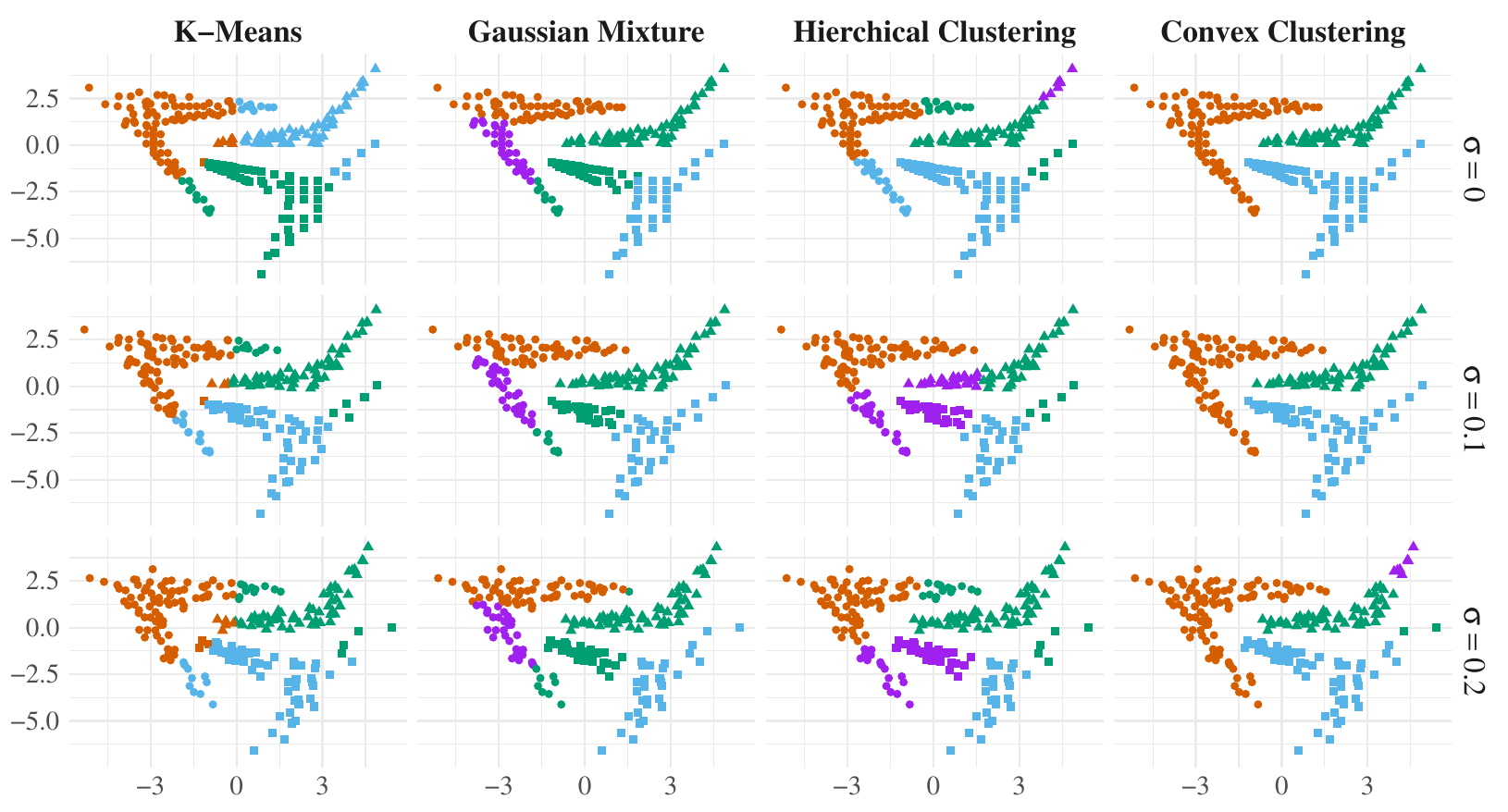}  
    \caption{Experiments on star-shaped dataset highlight differences in performance between $k$-means clustering, Gaussian mixture models, hierarchical clustering with average linkage, and convex clustering on non-convex datasets and varying noise levels on the rows.}
    \label{fig:comparisons_starshaped}  
\end{figure}

We next present a small case study involving data where samples from clusters are drawn from non-elliptical distributions. We also study what happens to the clustering results as noise is added to these samples.  Unlike the point clouds observed in the mammals dataset, Figure \ref{fig:comparisons_starshaped} depicts examples with a synthetic star-shaped dataset, which we revisit in greater detail later.  We utilize this synthetic dataset to illustrate non-elliptical data distributions for straight forward visualization.  The subpanels in Figure \ref{fig:comparisons_starshaped} show results with $k$-means clustering, GMMs, hierarchical clustering with average linkage, and convex clustering on the columns.  Along the rows, we add a small amount of entry-wise zero mean Gaussian noise with standard deviation $\sigma$ indicated on the right-side of the plots.  For completeness, experiments involving hierarchical clustering with average, single, complete, and Ward linkage can be found in Figure S1 in the Supplemental Materials.  We show only representative best results using average linkage in Figure \ref{fig:comparisons_starshaped}.

In each subplot of Figure \ref{fig:comparisons_starshaped}, the horizontal and vertical axes depict the first and second principal components of the dataset.  Colors indicate clustering assignments obtained by the methods indicated on the columns.  Shapes indicate the true clustering assignment labels.  For each method, we compute clustering results for fewer than $5$ clusters and employ the extended BIC criterion (eBIC) \citep{chen2008extended} to select the best results.  To compute the eBIC for $k$-means clustering, GMMs, and hierarchical clustering, we employ the clustering assignments to compute group centers as the mean of observations in each group.  We then feed the following centroid matrix into the eBIC computation: the matrix has the same dimensions as the original dataset but in lieu of its original entries, each observation takes on the values of its assigned group center.  The final selected cluster size for each method is indicated by the number of colors depicted in each subplot.

The top row in Figure \ref{fig:comparisons_starshaped} depicts results with no noise ($\sigma = 0$).  Here, we observe that convex clustering accurately recovers the three true star-shaped groups.  At the same time, we observe that $k$-means clustering and GMM struggle with the non-convexity of the data distribution, as indicated by difficulty in assignments on the boundaries.  Indeed, $k$-means clustering and GMMs perform better with ellipsoid-shaped data since they assume Gaussianity.  Similar to $k$-means and GMM, hierarchical clustering with average linkage also seems to struggle with non-convex data.  As seen in Figure S1 of the Supplemental Materials, hierarchical clustering with single linkage is less likely to group together a large fraction of points in disjoint clusters but performs even worse than average linkage in clustering this dataset by grouping together entire disjoint clusters.

The second and third rows in Figure \ref{fig:comparisons_starshaped} depict results with small amounts of additive noise ($\sigma = 0.1$ and $\sigma = 0.2$).  Here, we observe that hierarchical clustering can be highly unstable to small perturbations in the data: a small amount of noise can drastically alter the clustering assignments.  Meanwhile, $k$-means clustering and GMMs continue to struggle primarily with the non-convexity of the data distributions.  At $\sigma = 0.2$, we see that the convex clustering results are affected at points where the noise has resulted in low density point clouds. 

These simple examples illustrate that convex clustering has some distinct comparative advantages over more common clustering approaches. Indeed, as we will discuss later, convex clustering comes with stability guarantees as well as the provable ability to cluster data sampled from non-convex point cloud distributions, provided sufficiently dense sampling. The rest of the paper walks readers through details in the convex clustering approach that give rise to these advantages. Given space constraints, we focus our attention on giving intuition and insight into the behavior and provable guarantees for convex clustering -- the why -- as well as practical guidance -- the how. Moreover, we do not attempt to cover all algorithms for convex clustering, but focus on the most commonly used algorithms in the literature to clarify how the computational costs scale with problem size. In the interest of space, we regrettably cannot cover the many extensions and applications of convex clustering. Readers can find additional extensions, applications, and algorithms in survey papers by \cite{Feng2024} and \cite{yousif2024convex}. Nonetheless, our goal in our curated review of convex clustering is to equip the reader with understanding so that they might see for themselves new ways to productively incorporate convex clustering into their statistical modeling.

We begin our review by introducing the SON convex clustering formulation as well as some of its distinguishing properties which we will refer to in the rest of our article. 
Section 2 is devoted to an in-depth exploration of the set of hyperparameters in the convex clustering problem: the weights. Under what conditions are we guaranteed to recover a tree and how should one set them in practice? Section 3 is devoted to theoretical results. We cover conditions under which perfect recovery is possible, as well as statistical error bounds. Section 4 reviews algorithms at a high level. 
We see that weights again enter the picture; the number of nonzero weights directly impacts the computational costs and therefore scalability of algorithms. Sections 5 covers additional practical considerations. Section 6 covers extensions of convex clustering. 

\subsection{Problem Set up and Basic Properties}

Given $n$ points $\mathcal{X} = \{x_1,\ldots, x_n\} \subset \Real^p$, 
we seek cluster centers (centroids) $u_i$ in $\Real^p$ for each point $x_i$ that minimize a  convex criterion $E_\gamma(u)$ 
\begin{eqnarray}
\label{eq:objective_function}
\underset{u \in \Real^{np}}{\text{minimize}}\ E_{\gamma}(u) & := & \frac{1}{2}\sum_{i=1}^n \lVert x_i-u_i\rVert^2_2 + \gamma \sum_{(i,j) \in \E}w_{ij} \lVert u_i-u_j \rVert_2,
\end{eqnarray}
where $\gamma$ is a nonnegative tuning parameter, $w_{ij}$ is a nonnegative weight that quantifies the similarity between $x_i$ and $x_j$, and $u$ is the vector in $\Real^{np}$ obtained by stacking the vectors $u_1, \ldots, u_n$ on top of each other. The set $\E = \{(i,j) \subset [n] \times [n] : w_{ij} > 0\}$ is the index set of the positive weights, and $[n]$ denotes the index set $\{1, \ldots, n\}$.

The residual sum of squares term in the objective function of Problem \Eqn{objective_function} quantifies how well the centroids $u_i$ approximate the data $x_i$, while the regularization term penalizes the differences between pairs of centroids $u_i$ and $u_j$. Other norms, such as the 1-norm and infinity-norm, can be used \citep{PelDeSuy2005, Hocking2011}, but for simplicity we focus on the more commonly used 2-norm. While many of the theoretical results hold for arbitrary norms, some norms lead to simpler computations than others.  The regularization term is a composition of the group lasso \citep{YuanYi2006} and the fused lasso \citep{TibSauRos2005} and incentivizes sparsity in the pairwise differences of centroid pairs via the nondifferentiability of norms at the origin. This can be seen as a simpler and convex version of the Gaussian mixture model proposed by \cite{Guo2009}. Overall, $E_{\gamma}(u)$ can be interpreted as the energy of a configuration of centroids $u$ for a given relative weighting $\gamma$ between model fit and model complexity as quantified by the regularization term. It will sometimes be helpful to view the convex clustering problem as smoothing a multivariate signal over a graph $\mathcal{G} = ([n], \mathcal{E})$.

The centroids vector $u$ that solves Problem \Eqn{objective_function} is a function of $\gamma$, the data, and the weights. Most of the time, we write it as a function of solely $\gamma$ and suppress its dependency on the data and weights for notational ease. Because the objective function $E_{\gamma}$ in Problem \Eqn{objective_function} is strongly convex, for each value of $\gamma$, it possesses a unique minimizer $u(\gamma)$. We denote $u(\gamma)$'s $i$th subvector in $\Real^p$ corresponding to the $i$th optimal centroid by $u_i(\gamma)$. The tuning parameter $\gamma$ trades off the relative emphasis between data fit and differences between pairs of centroids. When $\gamma=0$, the minimum is attained when $u_i=x_i$, namely when each point occupies a unique cluster. As $\gamma$ increases, the regularization term encourages cluster centroids to fuse together. Two points $x_i$ and $x_j$ with $u_i=u_j$ are said to belong to the same cluster. For sufficiently large $\gamma$, say $\gamma^*$, the $u_i$ fuse into a single cluster, namely $u_i(\gamma^*) = \overline{x}$, where $\overline{x}$ is the average of the data $x_i$ \citep{Chi2015, Tan2015}. The unique global minimizer $u(\gamma)$ is a continuous function of $\gamma$ \citep{Chi2017}; we refer to the continuous paths $u_i(\gamma)$, traced out from each $x_i$ to $\overline{x}$ as $\gamma$ varies, collectively as the solution path. Thus, by computing $u_i(\gamma)$ for a sequence of $\gamma$ over an appropriately sampled range of values, we can potentially recover a hierarchical clustering of $\mathcal{X}$.

The minimizer to Problem \Eqn{objective_function} is actually smooth in not just $\gamma$ but in {\em all} inputs into the problem. Let $x$ be the vector in $\Real^{np}$ obtained by stacking the vectors $x_1, \ldots, x_n$ on top of each other.  The cluster centroids vector $u$ is jointly continuous over $(x, \gamma, \{w_{ij}\})$ \citep{Chi2017} and is 1-Lipschitz in the data $x$ \citep{Chi2018}. This latter property warrants further explanation. 
Fix the weights $\{w_{ij}\}$ and tuning parameter $\gamma$, and let $\tilde{u}(x) = u(\gamma, x, \{w_{ij}\})$ be the cluster centroids computed using the data $x$ and $\tilde{u}(x + \Delta x) = u(\gamma, x + \Delta x, \{w_{ij}\})$ be the cluster centroids computed using the perturbed data $x + \Delta x$. Then
\begin{eqnarray}
\label{eq:stability}
\lVert \tilde{u}(x) - \tilde{u}(x + \Delta x) \rVert_2 & \leq & \lVert \Delta x \rVert_2.
\end{eqnarray}
The above inequality says that $u$ is stable in the sense that a small perturbation $\Delta x$ in the data $x$ is guaranteed to {\em not} result in disproportionately wild variations in the output. In fact, the change in $u$ cannot exceed the change in the data $\Delta x$.

To conclude this section, we define some notation that we use in the upcoming sections. 
Let $\mathcal{P} = \{\mathcal{P}_1, \mathcal{P}_2, \ldots, \mathcal{P}_K\}$ denote a partition of $\mathcal{X}$, i.e., $\mathcal{P}_k \subset \mathcal{X}$ for $k \in [K]$, 
\begin{eqnarray*}
\mathcal{X} & = & \cup_{k = 1}^K \mathcal{P}_k \quad\quad \text{and} \quad\quad \mathcal{P}_i \cap \mathcal{P}_j = \emptyset \quad \text{for all $i \neq j$}.
\end{eqnarray*}
A partition $\mathcal{P}$ on $\mathcal{X}$ induces an index partition $\mathcal{I} = \{ \mathcal{I}_1, \ldots, \mathcal{I}_K\}$ where
$\mathcal{I}_k = \{i : x_i \in \mathcal{P}_k\} \subset [n]$ and $k \in [K]$. Let $n_k = \lvert \mathcal{P}_k \rvert$ denote the number of observations in the $k$th block of the partition $\mathcal{P}$. 

\section{WEIGHTS}
\label{sec:weights}

The weights $w_{ij}$ play a critical role in clustering quality both in theory and practice. Intuitively, $w_{ij}$ should be inversely proportional to the distance between $x_i$ and $x_j$. Then, the centroid $u_i(\gamma)$ is a nonlinear average of $\mathcal{X}$ with more weight given to points most similar to $x_i$. When $\gamma$ is small, the averaging is local. As $\gamma$ increases, the averaging continuously becomes more global. 

\subsection{Tree Recovery}
To understand why the solution path can recover a tree, and consequently a hierarchical clustering of the data, we first consider a scenario that leads to a ``flat" clustering of the data.  If $\mathcal{G}$ consists of $K$ connected components then there is an index  partition $\mathcal{I} = \{\mathcal{I}_1, \mathcal{I}_2, \ldots, \mathcal{I}_K\}$ of $[n]$ and a corresponding partition of the edges $\{\E_1, \ldots, \E_K\}$ where $\E_k = \{ (i,j) : i, j \in \mathcal{I}_k\}$ for $k \in [K]$. Then the objective function of Problem \Eqn{objective_function} can be expressed as
\begin{eqnarray}
\label{eq:objective_function_connected_components}
E_\gamma(u) & = & \sum_{k=1}^K E^k_\gamma(u),
\end{eqnarray}
where
\begin{eqnarray}
\label{eq:objective_function_single_component}
E^k_\gamma(u) & = & \frac{1}{2}\sum_{i \in \mathcal{I}_k} \lVert x_i-u_i\rVert^2_2 + \gamma \sum_{(i,j) \in \E_k}w_{ij} \lVert u_i-u_j \rVert_2.
\end{eqnarray}
Since the functions $E^k_\gamma$ depend on disjoint subsets of the centroids vector $u$, the objective function in Problem  \Eqn{objective_function_connected_components} can be minimized by minimizing the $K$ functions in Equation \Eqn{objective_function_single_component} separately. Moreover, $u_i(\gamma) = \frac{1}{\lvert \mathcal{I}_k \rvert} \sum_{i \in \mathcal{I}_k} x_i$ for all $i \in \mathcal{I}_k$ when $\gamma$ is  sufficiently large \citep{PelDeSuy2005, Chi2015}. 

Consider the situation where $K = 2$ and add edges between the two connected components  so that the objective function is 
\begin{eqnarray}
\label{eq:twocomponents}
E_\gamma(u) & = & E^1_\gamma(u) + E^2_\gamma(u) + \gamma \sum_{(i,j) \in \E_{12}} w_{ij}\lVert u_i - u_j \rVert_2,
\end{eqnarray}
where $\E_{12}$ is the set of edges between $\mathcal{I}_1$ and $\mathcal{I}_2$. Suppose  
that weights connecting pairs of points within $\mathcal{I}_1$ or $\mathcal{I}_2$ are much greater than weights connecting points across the two sets, i.e., $\max_{(i,j) \in \E_{12}} w_{ij} \ll \min_{(i,j) \in \E \backslash \E_{12}} w_{ij}$.
Then the objective function \Eqn{twocomponents} is nearly separable over the sets $\mathcal{I}_1$ and $\mathcal{I}_2$; the sum over $\E_{12}$ loosely couples $E^1_\gamma(u)$ and $E^2_\gamma(u)$ together. Since the solution path $u(\gamma)$ is continuous in the weights, we  expect that as $\gamma$ increases, local averaging of the centroids will first occur over $\mathcal{I}_1$ and $\mathcal{I}_2$ separately, i.e., the $\mathcal{I}_1$ centroids will fuse together into a common $\mathcal{I}_1$-centroid
and the $\mathcal{I}_2$ centroids will fuse together into a common $\mathcal{I}_2$-centroid. As $\gamma$ continues to increase the $\mathcal{I}_1$-centroid and $\mathcal{I}_2$-centroid will eventually fuse. This is the intuition behind the tree recovery property of convex clustering given in \cite{ChiSteinerberger2019}. They prove that a sufficient condition to ensure the solution path produces a tree that reflects the geometry of $\mathcal{X}$ is that the weights decay geometrically fast in the distance between points. The rate constant is data dependent.


\begin{figure}[htbp]
    \centering
    \begin{tabular}{*{3}{>{\centering\arraybackslash}m{0.3\textwidth}}}  
        
        \includegraphics[width=\linewidth]{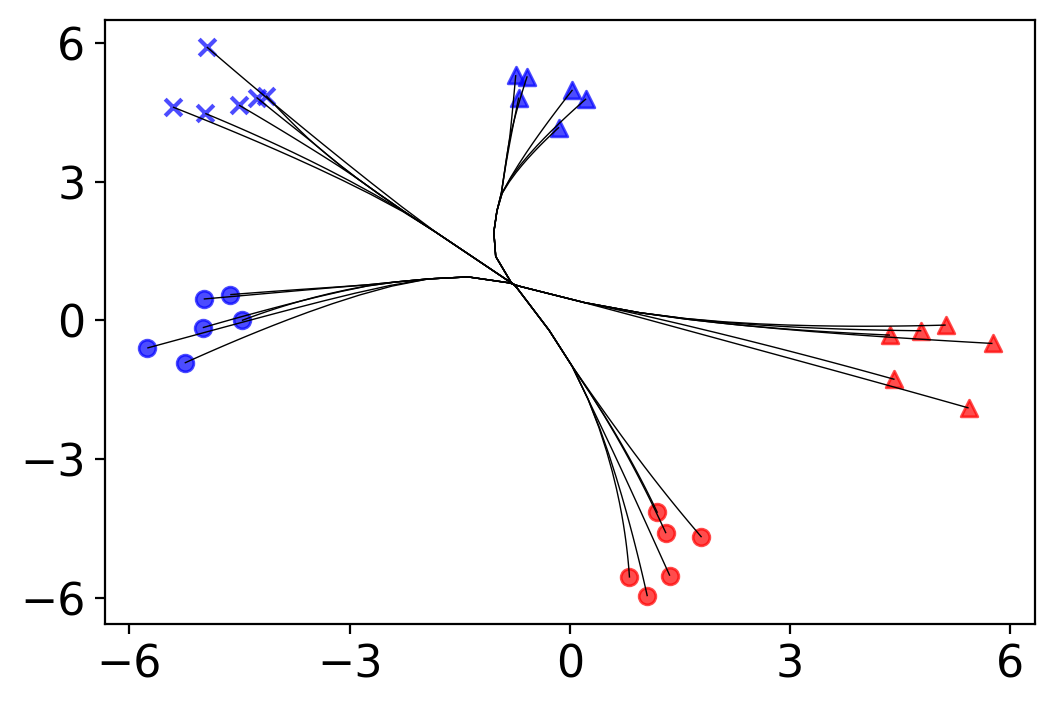} & \includegraphics[width=\linewidth]{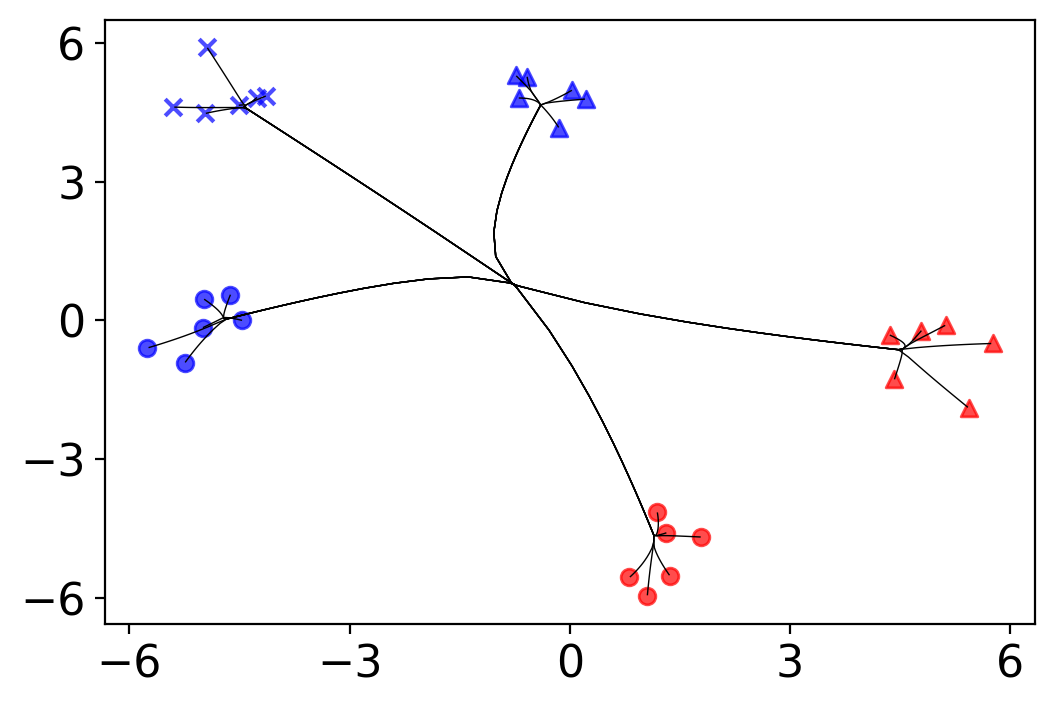} & \includegraphics[width=\linewidth]{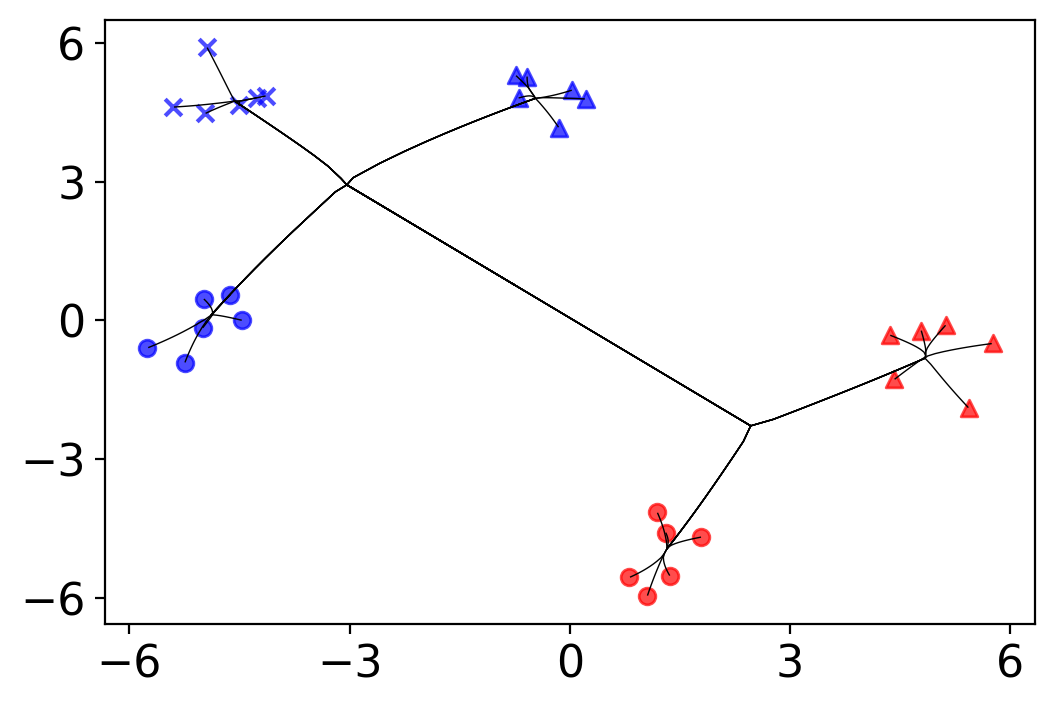} \\
        
        \\
    \end{tabular}
        
    \caption{Clustering paths with weights taking on one of three values: $w_{L_1} \geq w_{L_2} \geq w_{L_3}$. Subplots from left to right: $(w_{L_1}, w_{L_2}, w_{L_3})$  are $(1, 1, 1)$, $(10, 1, 1)$ and $(10, 1, 0.1)$.}\label{fig:level_weights}
\end{figure}





\Fig{level_weights} shows this sufficient condition in action. There are five clusters consisting of five points each. Points belonging to the same cluster share a common symbol, or color and shape combination. The five clusters can be further collected into two super-clusters based on proximity -- colored in the upper left in blue and the lower right in red. Thus, there are three levels of organization or hierarchy among the data. All weights are positive and take one of three values: $w_{L_1} \geq w_{L_2} \geq w_{L_3}$. Within cluster weights have the value $w_{L_1}$. Weights between points in different clusters, but within super-clusters   (points in the same color) have the value $w_{L_2}$. Weights between points in different super-clusters have the value $w_{L_3}$. In the left panel, when the weights are uniform, i.e., $w_{L_1} = w_{L_2} = w_{L_3}$, the solution path exhibits exactly {\em one} fusion event as $\gamma$ increases. In the middle panel, when the within cluster weights are an order of magnitude larger than all the other weights, the solution path exhibits fusions that initially group together centroids in their respective clusters. As $\gamma$ continues to increase, these cluster centroids remain distinct until they all simultaneously fuse at the end of the solution path. In the right panel, the weights across different levels of organization differ by an order of magnitude.  Consequently, as $\gamma$ increases, the centroids at the cluster level fuse first and then fuse at the super-cluster level before, finally, the two super-cluster centroids fuse together. 

\subsection{Weights in Practice}
We next discuss choosing weights in practice. Note that it is possible to choose weights so that the solution path is not a tree. \cite{Hocking2011} provide a simple example of this. Fortunately, generating this degeneracy requires adversarially chosen weights that are inconsistent with the geometry of the data, i.e., weights that are not inversely proportional to the distances between points. With this in mind, a natural choice is the inverse Euclidean distance between points, i.e.,
\begin{eqnarray*}
w_{ij} & = & \frac{1}{\lVert x_i - x_j \rVert_2}.
\end{eqnarray*}
Perhaps the most commonly used weight, however, is the Gaussian kernel weight
\begin{eqnarray*}
    w_{ij} & = & \exp\left(-\frac{\lVert x_i - x_j \rVert_2^2}{\sigma_{ij}}\right).
\end{eqnarray*}
Gaussian kernel weights decay faster with distance than the sufficient condition in \cite{ChiSteinerberger2019}, which explains why they work well in practice. The scale parameter $\sigma_{ij}$ should be data dependent. \cite{Zelnik-Manor2005} propose a measure of local scale for spectral clustering that directly applies. They prescribe setting $\sigma_{ij} = \sigma_i\sigma_j$, where $\sigma_i$ is a local measure of scatter around $x_i$, e.g., the median Euclidean distance between the $x_i$ and its nearest neighbors. In the examples below, we use  $\left\lfloor \frac{n}{10} \right\rfloor$  neighbors, but a data-independent number of nearest neighbors can also be used, e.g., 3 to 7 nearest neighbors.

It is also important for the graph $\mathcal{G}$ to be sparse, i.e., $\lvert \E \rvert \ll n^2$. Indeed, an oracle would set $w_{ij} > 0$ if and only if $x_i$ and $x_j$ belonged to the same cluster. Selective inclusion of edges in $\E$ can significantly improve clustering quality. Convex clustering can again leverage design decisions from spectral clustering, which also benefits from sparse affinity graphs, to construct a sparse $\E$. Table~\ref{tab:graph_construction} lists some ways to construct $\E$ such that $\lvert \E \rvert = \mathcal{O}(n)$. See \cite{ZemelCarreira2004} for more discussion on methods for constructing $\E$, including their proposed disjoint minimum spanning trees (\texttt{DMST}s).




\begin{table}[htbp]
    \centering
    \caption{Graph Construction Methods}   
    \label{tab:graph_construction}
    \begin{tabular}{lp{10cm}} 
        \hline
        \textbf{Method} & \textbf{Description} \\ 
        \hline
        \texttt{MST} & Minimum spanning tree \\ 
        $k$-\texttt{NNG} & $k$-nearest neighbors graph \\ 
        \texttt{MST} + $k$-\texttt{NNG} & Union of \texttt{MST} and $k$-\texttt{NNG} \\ 
        \texttt{DMSTs} & A graph that is a collection of $M$ \texttt{MST}s where the $m$th tree for $m \in [M]$ is the \texttt{MST} of the data subject to not using any edge already in the previous $1, \dots, M-1$ trees. \\ 
        \hline
    \end{tabular}
\end{table}

\begin{figure}[htbp]
    \centering

    \begin{tabular}{>{\centering\arraybackslash}m{0.08\textwidth} *{4}{m{0.20\textwidth}}}  
         
        & \makecell{Graphs}  & \makecell{Uniform \\ weights} & \makecell{Inverse Euclidean \\ weights} & \makecell{Gaussian kernel \\ weights} \\
        
        \texttt{MST} &
        \includegraphics[width=\linewidth]{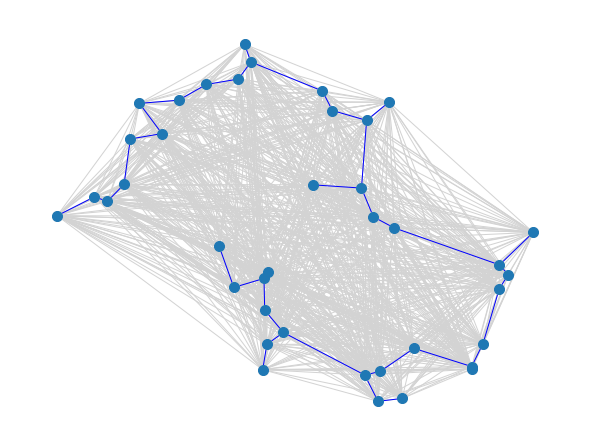} &
        \includegraphics[width=\linewidth]{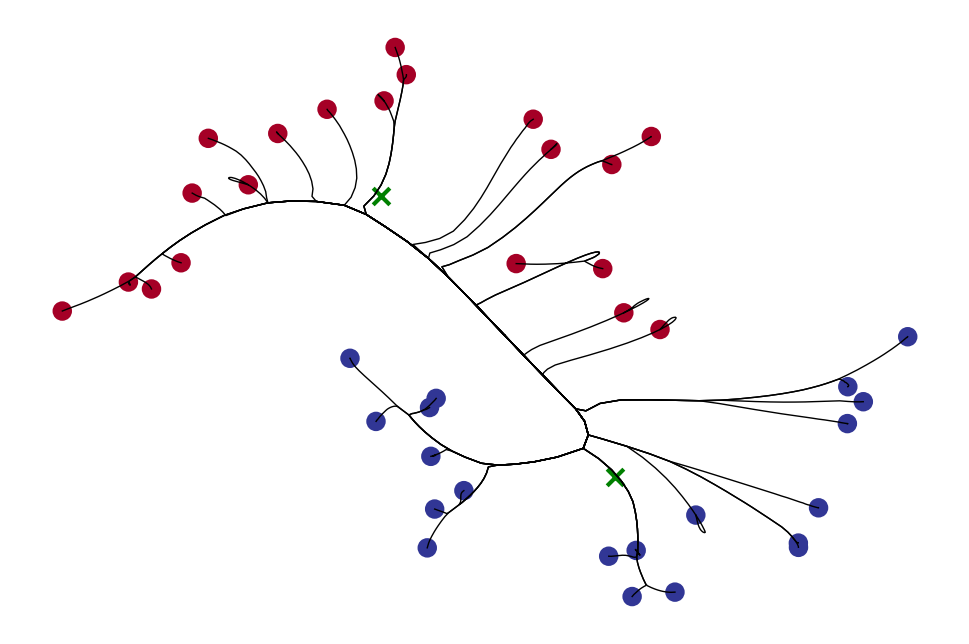} &
        \includegraphics[width=\linewidth]{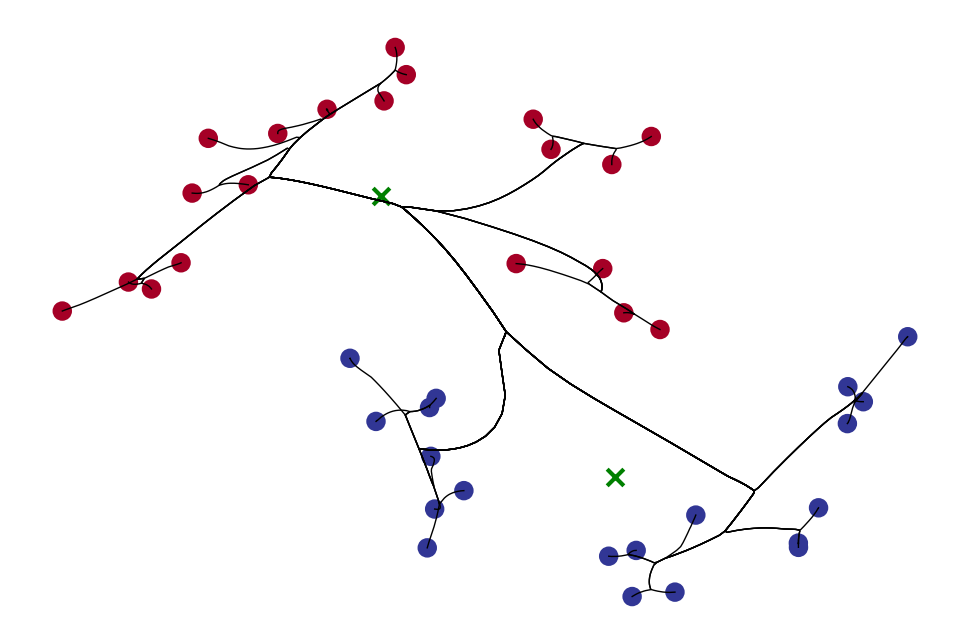} &
        \includegraphics[width=\linewidth]{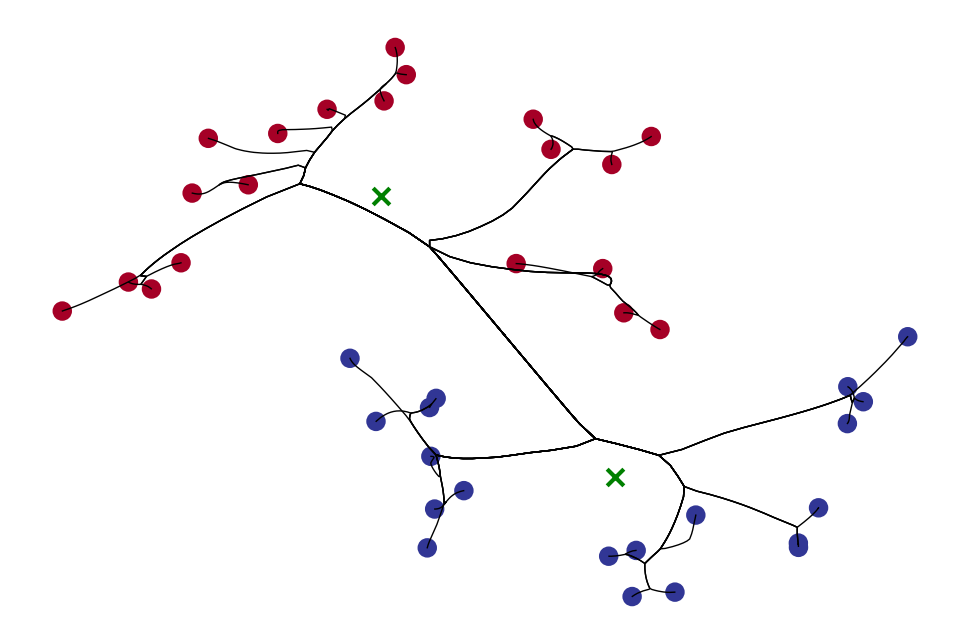} \\
        
        \texttt{$3$-NNG} &
        \includegraphics[width=\linewidth]{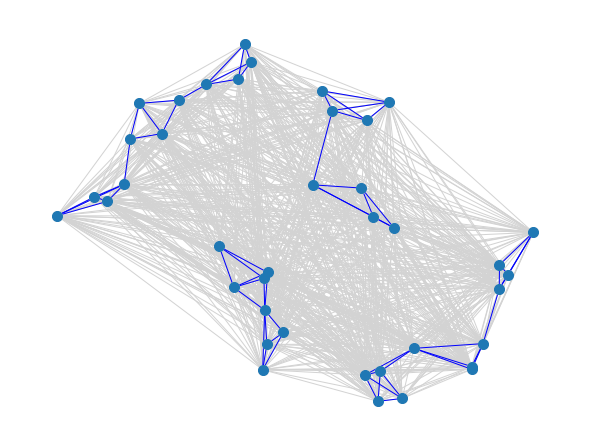} &
        \includegraphics[width=\linewidth]{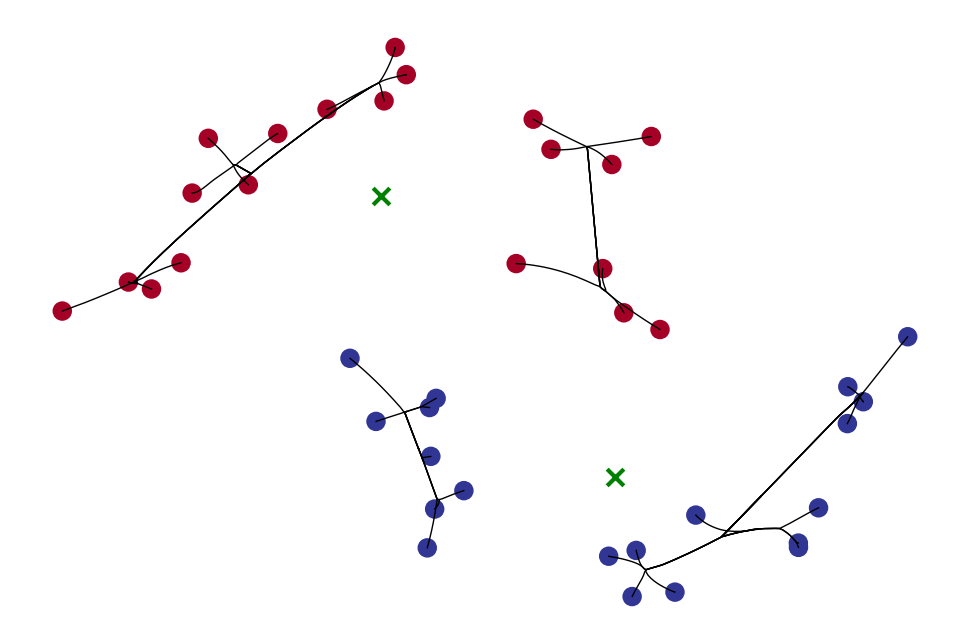} &
        \includegraphics[width=\linewidth]{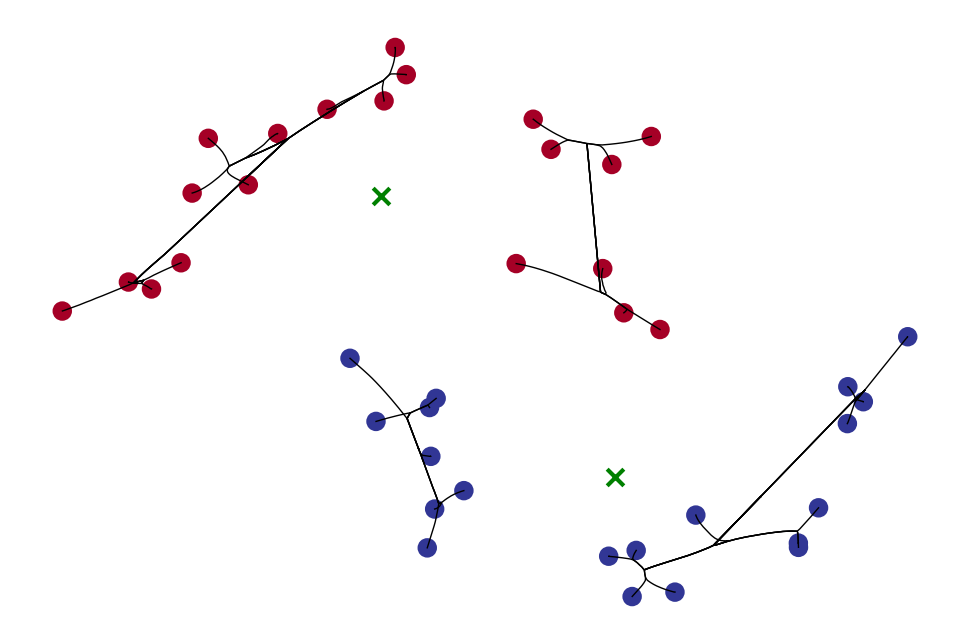} &
        \includegraphics[width=\linewidth]{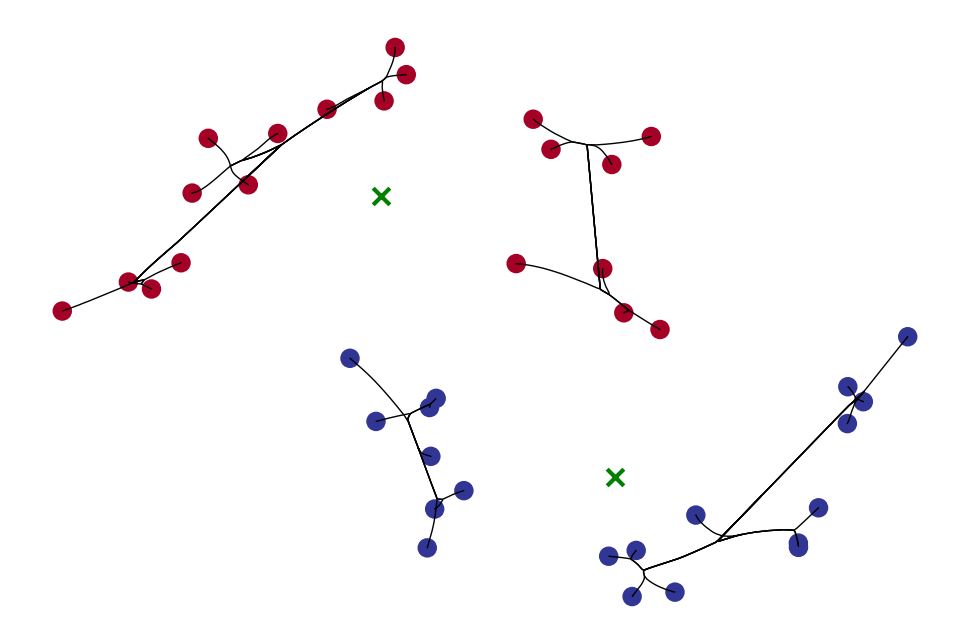} \\

        \makecell{\texttt{$3$-NNG} \\ + \texttt{MST}} &
        \includegraphics[width=\linewidth]{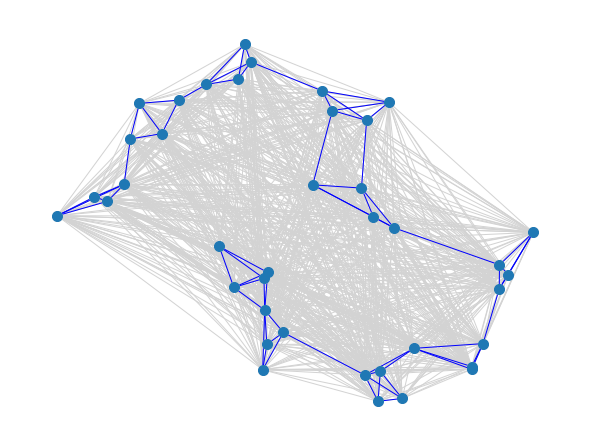} &
        \includegraphics[width=\linewidth]{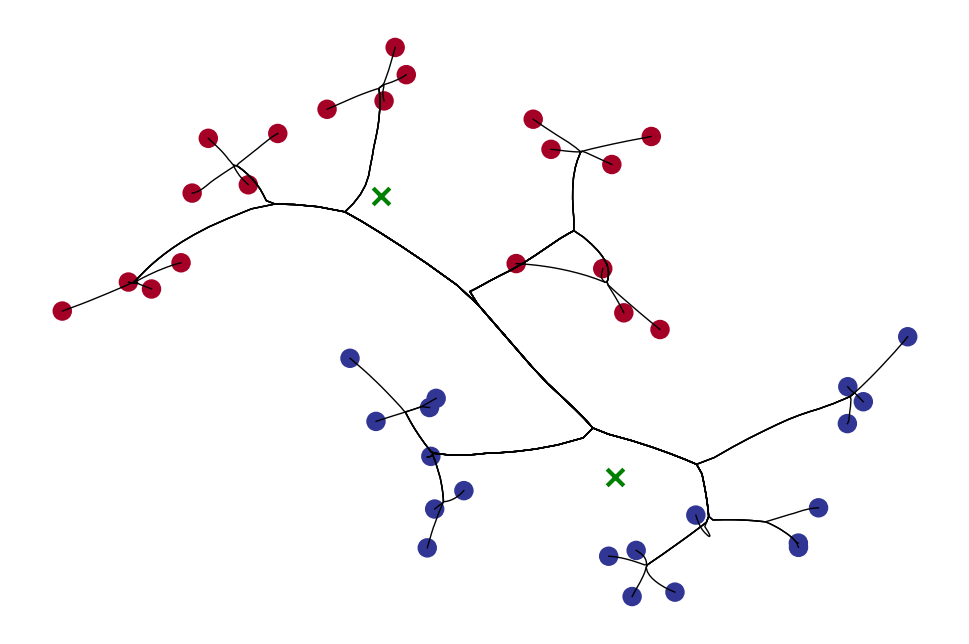} &
        \includegraphics[width=\linewidth]{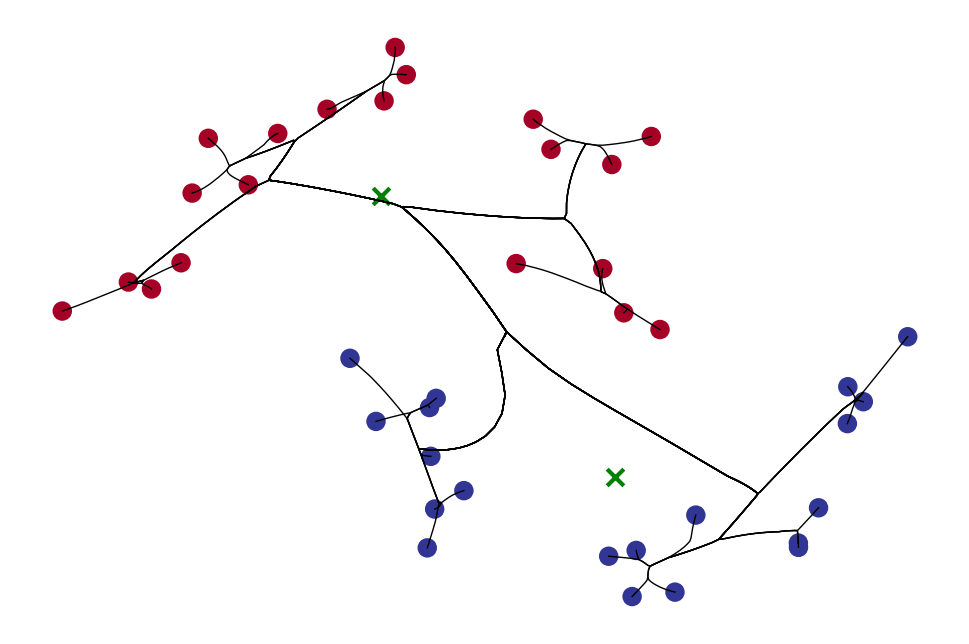} &
        \includegraphics[width=\linewidth]{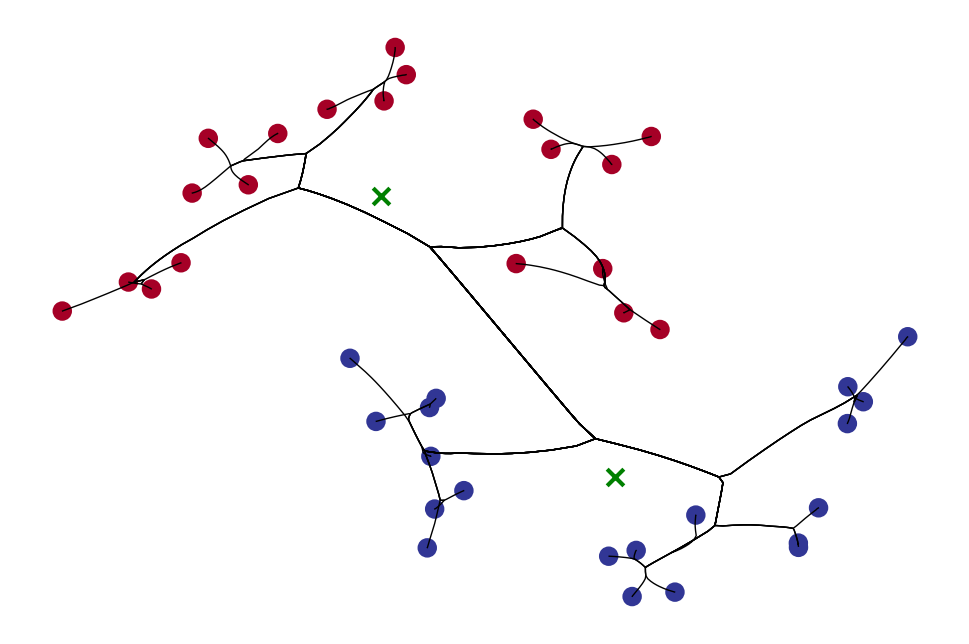} \\

        \makecell{\texttt{DMSTs} \\ ($M = 3$)} &
        \includegraphics[width=\linewidth]{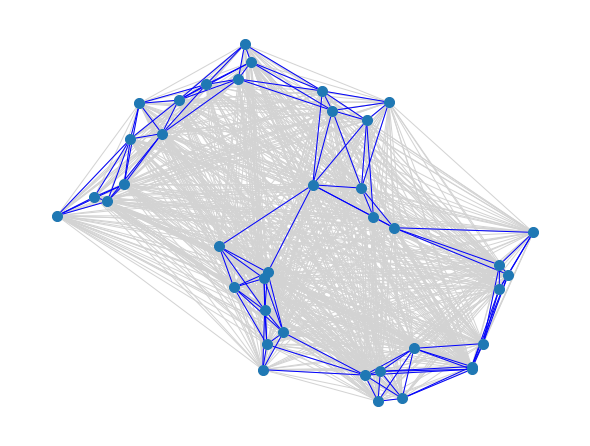} &
        \includegraphics[width=\linewidth]{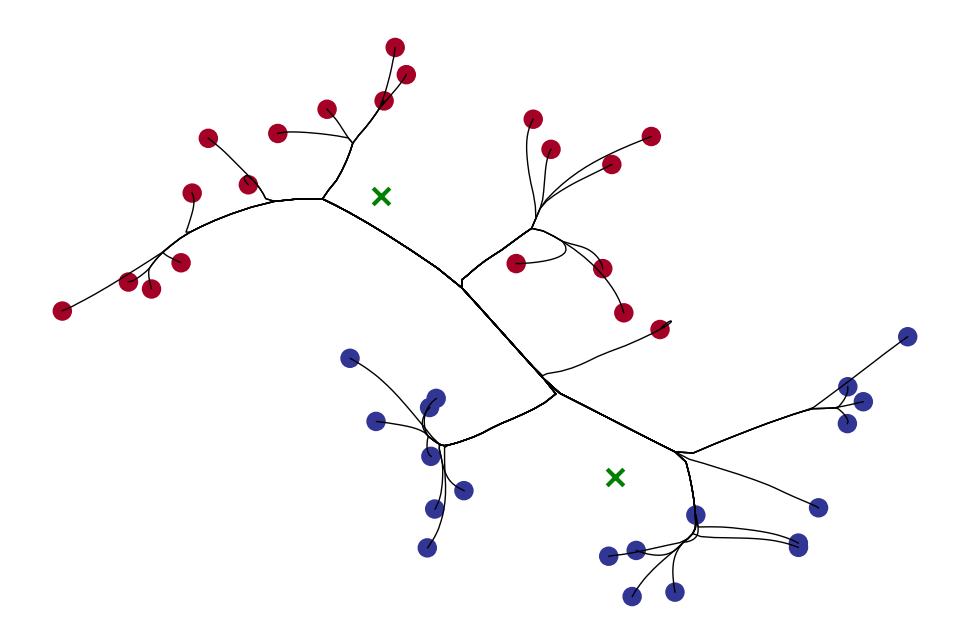} &
        \includegraphics[width=\linewidth]{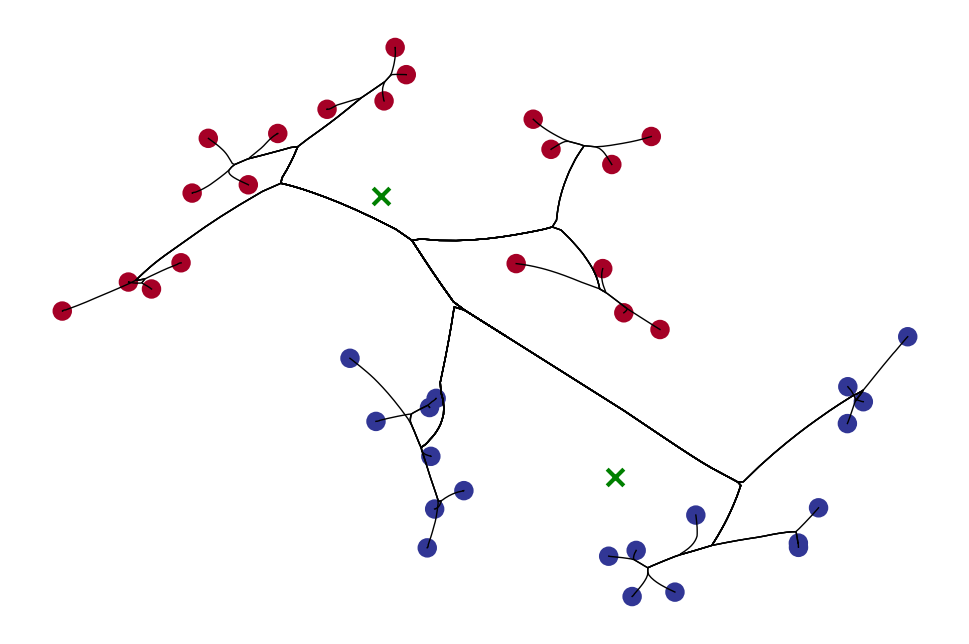} &
        \includegraphics[width=\linewidth]{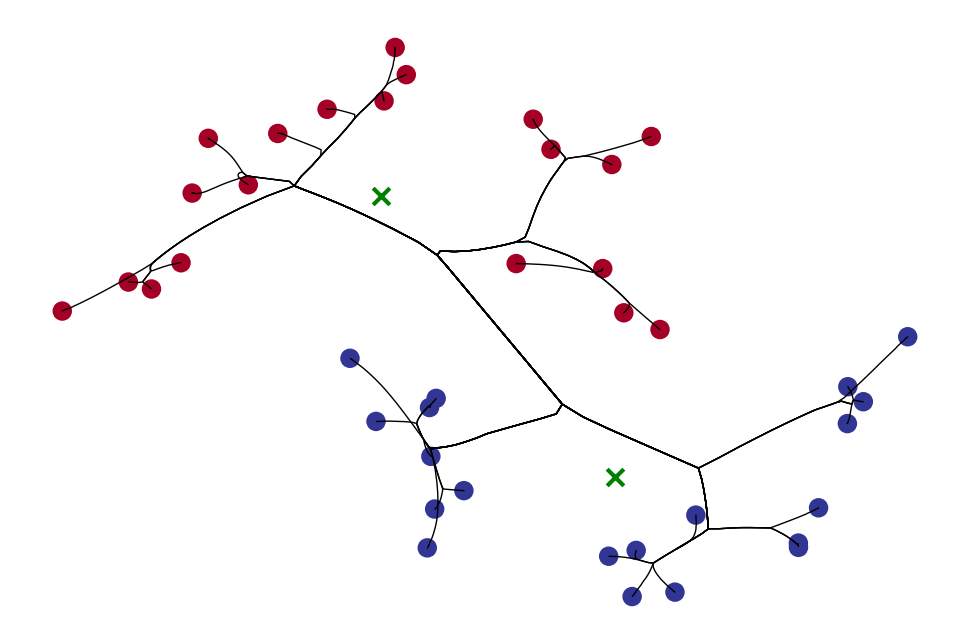} \\
        
    \end{tabular}

    \caption{A comparison of different choices of graph construction and weight choices along with solution paths on a pair of half-moons dataset. The dataset is created with 20 points in the red and blue clusters, respectively. Th first column contains graphs constructed by \texttt{MST}, $3$-\texttt{NNG}, \texttt{MST} + $3$-\texttt{NNG} and \texttt{DMST}s $(M = 3)$. The second, third and fourth columns contain corresponding solution paths using uniform weights, inverse Euclidean weights, and Gaussian kernel weights on the different graphs. The green x's are the sample means of the two half moons}\label{fig:comparison-of-solution-paths-with-different-graph-construction-and-weights}
\end{figure}


To illustrate the effect of different combinations of graph construction methods and weight types, we apply convex clustering on simulated data of two interlocking half moons. \Fig{comparison-of-solution-paths-with-different-graph-construction-and-weights} shows the resulting solution paths. Not surprisingly, to obtain a tree structure, $\mathcal{G}$ needs to be connected. Additionally, the uniform weights tend to recover the worst solution paths. Interestingly, the inverse Euclidean weights recover  similar trees to the Gaussian kernel weights even though they do not satisfy the sufficient conditions in \cite{ChiSteinerberger2019},  furnishing evidence that they are sufficient but not necessary conditions for tree recovery.

\subsection{Connection to Clustering with Non-convex Fusion Penalties}
Weights provide a connection between convex clustering and other penalized regression-based clustering methods that use folded-concave penalties \citep{PanShenLiu2013, Marchetti2014,Wu2016}. These methods seek to minimize the following variation on objective function of Problem \Eqn{objective_function}
\begin{eqnarray}
\label{eq:folded_concave_objective}
E^\varphi_{\gamma}(u) & = & \frac{1}{2}\sum_{i=1}^n \lVert x_i-u_i\rVert^2_2 + \gamma \sum_{(i,j) \in \E}\varphi_{ij}\left(\lVert u_i-u_j \rVert_2\right),
\end{eqnarray}
where each function $\varphi_{ij} : [0,\infty) \mapsto [0, \infty )$ has the following properties: (i) $\varphi_{ij}$ is concave and differentiable on $(0, \infty)$, (ii) $\varphi_{ij}$ vanishes at the origin, and (iii) the directional derivative of $\varphi_{ij}$ exists and is positive at the origin. The penalty function $\varphi_{ij}$ is called a folded-concave penalty. Important examples include the smoothly clipped absolute deviation \citep{FanLi2001} and minimax concave penalty \citep{Zha2010}. The folded concave fusion penalties in \Eqn{folded_concave_objective} have been effectively used in a variety of statistical models that incorporate simultaneous clustering, e.g., \cite{ma2017concave}, \cite{zhu2018cluster}, \cite{austin2020new}, and \cite{zhu2021longitudinal}.

It is well known that nonsmooth convex penalties suffer from shrinkage bias. In the context of convex clustering, the centroid estimates $u_i$ are biased towards the grand mean $\overline{x}$. Folded concave penalties suﬀer far less bias in exchange for giving up convexity in the optimization problem. This exchange means that iterative algorithms are typically guaranteed to converge only to a KKT point at best.

Since $\varphi_{ij}$ in \eqref{eq:folded_concave_objective} is concave and differentiable, for all positive $z$ and $\tilde{z}$
\begin{eqnarray}
\label{eq:folded_concave_Taylor}
	\varphi_{ij}(z) & \leq & \varphi_{ij}(\tilde{z}) + \varphi'_{ij}(\tilde{z})(z - \tilde{z}).
\end{eqnarray}
The inequality \Eqn{folded_concave_Taylor} indicates that the first order Taylor expansion of a differentiable concave function $\varphi_{ij}$ provides a tight global upper bound at the expansion point $\tilde{z}$. Thus, we can construct a function that is a tight upper bound of the function $\tilde{E}_\gamma(u)$
\begin{eqnarray*}
g_\gamma(u \mid \tilde{u}) & = & 
\frac{1}{2}\sum_{i=1}^n \lVert x_i-u_i\rVert^2_2 + \gamma \sum_{(i,j) \in \E}\tilde{w}_{ij} \lVert u_i - u_j \rVert_2 + c,
\end{eqnarray*}
where the constant $c$ does not depend on $u$ and $\tilde{w}_{ij}$ are weights that depend on $\tilde{u}$, i.e.,
\begin{eqnarray*}
\label{eq:majorization}
\tilde{w}_{ij} & = & \varphi'_{ij}\left(\lVert \tilde{u}_i - \tilde{u}_j \rVert_2 \right).
\end{eqnarray*}
Notice that in the framework of \eqref{eq:folded_concave_objective}, we recover Gaussian kernel weights by taking
\begin{eqnarray*}
\varphi_{ij}(z) & \propto & \int_{0}^z \exp\left(-\frac{\vartheta^2}{\sigma_i\sigma_j}\right) d\vartheta.
\end{eqnarray*}
Additionally, we recover approximate inverse Euclidean weights by taking $\varphi_{ij}(z) = \log(z + \delta)$ for a small positive $\delta$. Note that $\varphi(z) = \log(z)$ recovers inverse Euclidean weights exactly.  However, this does not satisfy properties (ii) and (iii) of folded concave penalties whereas $\log(z + \delta)$ does.

The function $g_\gamma(u \mid \tilde{u})$ is said to majorize the function $\tilde{E}_\gamma(u)$ at the point $\tilde{u}$  \citep{LanHunYan2000} and minimizing it corresponds to performing one step of the local linear-approximation algorithm \citep{Zou2008, Schifano2010}, which is an instance of a majorization-minimization (MM) algorithm \citep{LanHunYan2000}.  \cite{Zou2008} show that the solution to the one-step algorithm is often sufficient in terms of its statistical estimation accuracy.

\section{THEORETICAL RESULTS}
Theoretical results for convex clustering focus mainly on conditions for perfect recovery of the partition $\mathcal{P}$ and estimation of cluster centroids.  We discuss these results in separate subsections, since they often make fundamentally different assumptions. 

\subsection{Perfect Recovery}
Over the previous decade, much work has focused on sufficient conditions for recovering the partition $\mathcal{P}$. Following \citet{Panahi2017}, define the diameter of the set $\mathcal{P}_k$ as
\begin{eqnarray*}
D(\mathcal{P}_k) & = & \sup \{\|x - y\|_2: x, y \in \mathcal{P}_k\},
\end{eqnarray*}
and the distance between $\mathcal{P}_k$ and $\mathcal{P}_{k'}$ as
\begin{eqnarray*}
d(\mathcal{P}_k, \mathcal{P}_{k'}) & = & \inf \left\{\lVert x - y\rVert_2: x \in \mathcal{P}_k, y \in \mathcal{P}_{k'}\right\}.
\end{eqnarray*}
When $d$ is evaluated at two points in $\Real^p$, we define $d(x,y) = \|x - y\|_2$. 
Recall that $\mathcal{I}_k = \{i : x_i \in \mathcal{P}_k\}$ and $n_k = |\mathcal{I}_k|$, and define $\overline{x}_k = \frac{1}{n_k} \sum_{i \in \mathcal{I}_k} x_i$ for $k \in [K].$ Loosely speaking, perfect recovery is guaranteed when the smallest distance between clusters is sufficiently larger than the largest cluster diameter.

\citet{zhu2014convex} established the first known sufficient conditions for perfect recovery when all weights are positive and equal, i.e., uniform weights, 
under the following \textit{two-cubes} model. With $K = 2$, they 
assume that $\mathcal{P}_k$ has centroid $c_k = (c_{(k)1}, \dots, c_{(k)p})\Tra \in \Real^p$ and edge lengths $s_k = 2(s_{(k)1}, \dots, s_{(k)p})\Tra \in \Real^p,$ i.e., $x_i \in \mathcal{P}_k$ if and only if $x_i \in [c_{(k)1} - s_{(k)1}, c_{(k)1} + s_{(k)1}]\times \cdots \times [c_{(k)p} - s_{(k)p}, c_{(k)p} + s_{(k)p}].$ 
They further assume that $n_1$ points are drawn from the first cube (randomly or deterministically), and $n_2 = n - n_1$ points from the second. They define 
\begin{eqnarray*}
{\rm size}(\mathcal{X}, \mathcal{P}) & = & \max\{(2n_2(n_1 - 1)/n_1^2 + 1)\|s_{1}\|_2,(2n_1(n_2 - 1)/n_2^2 + 1)\|s_{2}\|_2\}.
\end{eqnarray*}
With these definitions, they prove that perfect recovery occurs if
\begin{eqnarray*}
\frac{2}{n}\,{\rm size}(\mathcal{X}, \mathcal{P}) & \leq & \gamma \amp \leq \amp \frac{2}{n}\,d(\mathcal{P}_1, \mathcal{P}_{2}).
\end{eqnarray*}
Note that the condition is vacuous if ${\rm size}(\mathcal{X}, \mathcal{P}) > d(\mathcal{P}_1, \mathcal{P}_2)$. These sufficient conditions are intuitive. First, if the cubes are too close together, or if the sample-weighted size of either cube is too large relative to its distance, perfect recovery may not occur. Second, ${\rm size}(\mathcal{X}, \mathcal{P})$ is minimized when the two clusters have the same number of points and grows  with the degree of imbalance between $n_1$ and $n_2$. A larger ${\rm size}(\mathcal{X}, \mathcal{P})$ requires  greater distance between $\mathcal{P}_1$ and $\mathcal{P}_2$ for the conditions to hold. If too many points are drawn from one cluster, a point in the smaller cluster will have more edges connecting it to the larger cluster than edges connecting it to its fellow smaller cluster points. Consequently, as $\gamma$ increases, its centroid will be pulled more strongly toward centroids in the larger cluster than toward  fellow smaller cluster centroids.

\citet{Panahi2017} extend the results of \citet{zhu2014convex} to more general partitions. For arbitrary $K$, \citet{Panahi2017} prove that perfect recovery occurs if $\gamma \in [\gamma_L, \gamma_U]$ where
\begin{eqnarray*}
\gamma_L & = & \underset{k \in [K]}{\max}\,\left \{\frac{D(\mathcal{P}_k)}{n_k} \right\}, \quad \text{and}\quad \gamma_U \amp = \amp \min_{k \neq {k'}}\, \left\{\frac{d(\bar{x}_k,\bar{x}_{k'})}{2n}\right\}.
\end{eqnarray*}
Like \citet{zhu2014convex}, the sufficient conditions of \citet{Panahi2017} require that the weighted diameters be sufficiently smaller than the distance between clusters quantified as the distance between their centroids. 
\citet{Panahi2017} also assume uniform weights and recover the result of \citet{zhu2014convex} as a special case. 

The perfect recovery conditions of \citet{zhu2014convex} and \citet{Panahi2017} are not probabilistic. Rather, their proofs establish sufficient conditions under which the optimality conditions for $u(\gamma) = \arg\min_u E_\gamma(u)$ are satisfied by $u(\gamma)$ with partition $\mathcal{P}.$ The results of \citet{Panahi2017}, however, can be applied to points generated from an isotropic Gaussian mixture model. Let $\mu_k \in \Real^{p}$ and $\sigma_k^2 I_p \in \mathbb{S}^p_+$ denote the mean and covariance for the $k$th Gaussian mixture component. Then the results of \citet{Panahi2017} imply that perfect cluster recovery occurs with high probability, i.e., correct assignment to the cluster defined by the generating mixing component, is possible if
\begin{eqnarray*}
 \frac{\max_v \sigma_v}{\min_v \pi_v} {\rm polylog}(n) & \lesssim & \min_{k \neq k'}d\left(\mu_k,\mu_{k'}\right) 
\end{eqnarray*}
for positive mixture probabilities $\pi_1, \dots, \pi_K$ \citep{jiang2020recovery}. The  ${\rm polylog}(n)$ term captures the following intuition. Consider a ball $B_k$ of arbitrary radius around the $k$th component mean $\mu_k$. As $n \to \infty$, the probability of observing a sample $x_i$ from the $\tilde{k}$th component, where $\tilde{k} \neq k$, land in $B_k$ tends to one unless the distances between component means diverge. Motivated by this observation, \citet{jiang2020recovery} establish i) a lower bound on $\gamma$ that ensures, with high probability, all points within a $\theta$-standard deviation neighborhood of $\mu_k$, i.e.,
\begin{eqnarray*}
A_k(\theta) & = & \left\{i: d\left(x_i,\mu_k\right) \leq \theta \sigma_k\right\},
\end{eqnarray*}
form one cluster, and ii) an upper bound that ensures $A_k$ and $A_{k'}$ remain distinct. As might be expected, the lower bound depends on $\theta, \sigma_k, $ and $n$ --- parameters that influence the sample size-weighted diameter of $A_k(\theta)$ --- whereas the upper bound depends on $d(\mu_k, \mu_{k'})/n$. The results in \citet{jiang2020recovery} do not focus on perfect recovery per se. Instead, they verify that under the Gaussian mixture generative model, convex clustering tends to assign points to clusters like a Gaussian mixture model would.

\begin{figure}[htbp]
    \centering
    \begin{tabular}{*{3}{>{\centering\arraybackslash}m{0.3\textwidth}}}  
        
        \includegraphics[width=\linewidth]{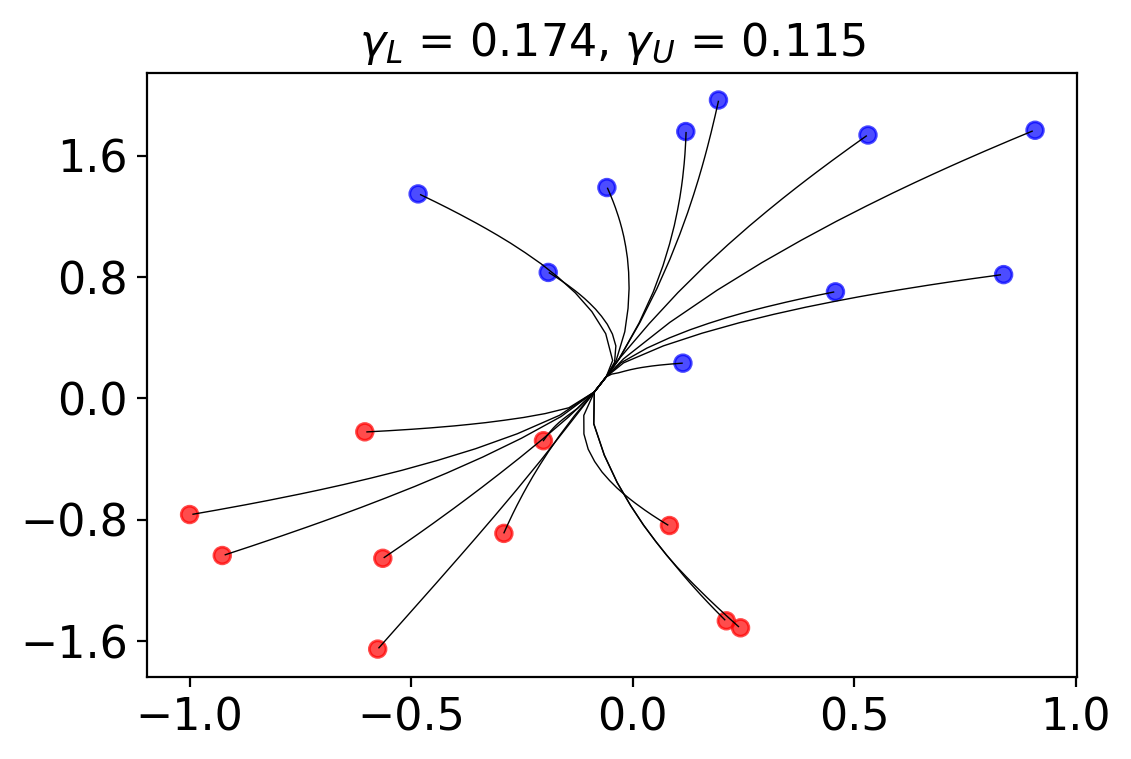} & \includegraphics[width=\linewidth]{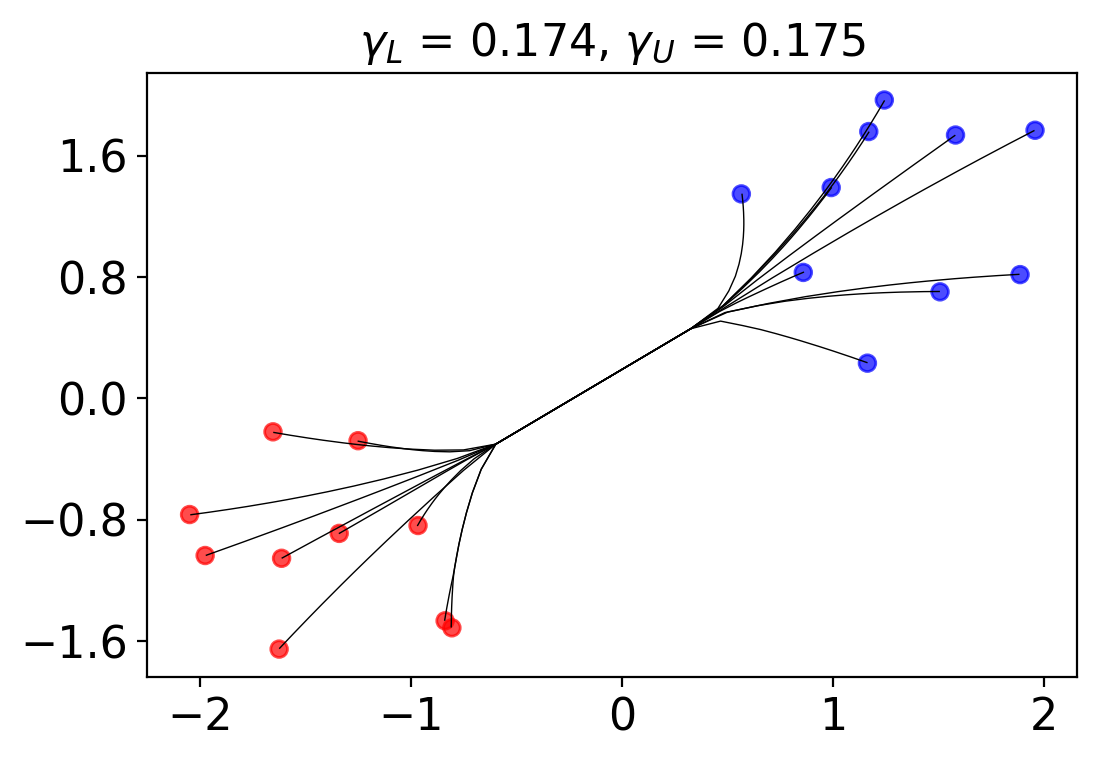} & \includegraphics[width=\linewidth]{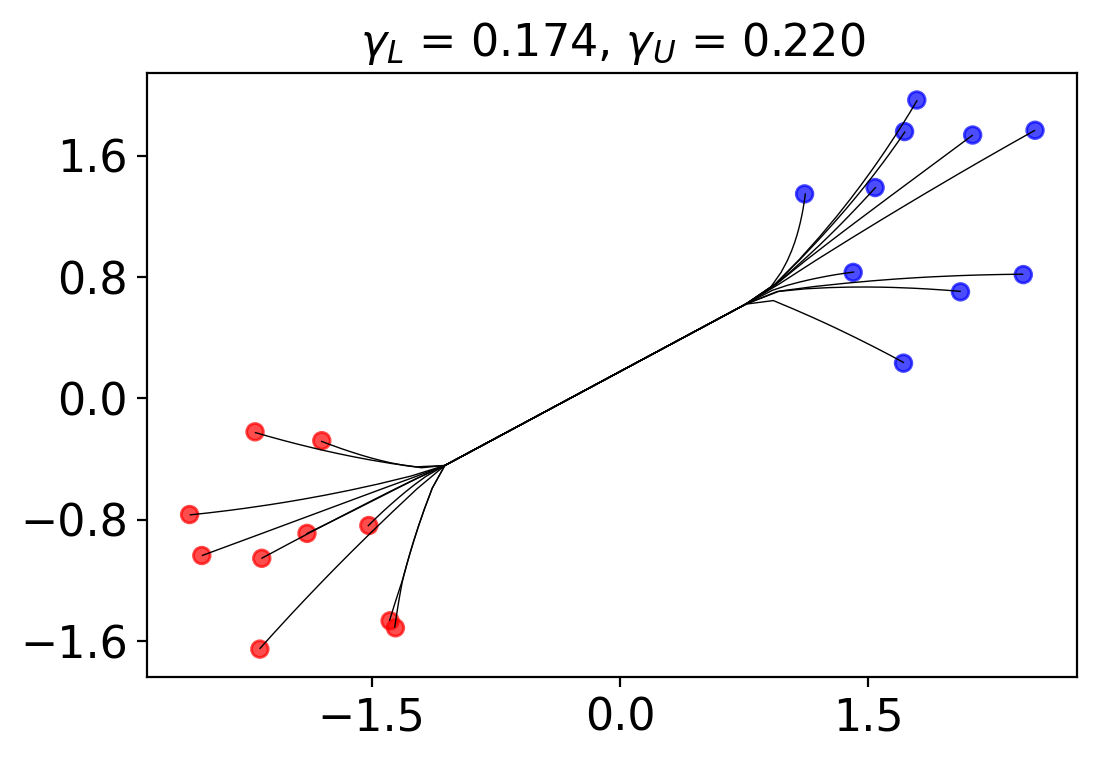} \\
        \includegraphics[width=\linewidth]{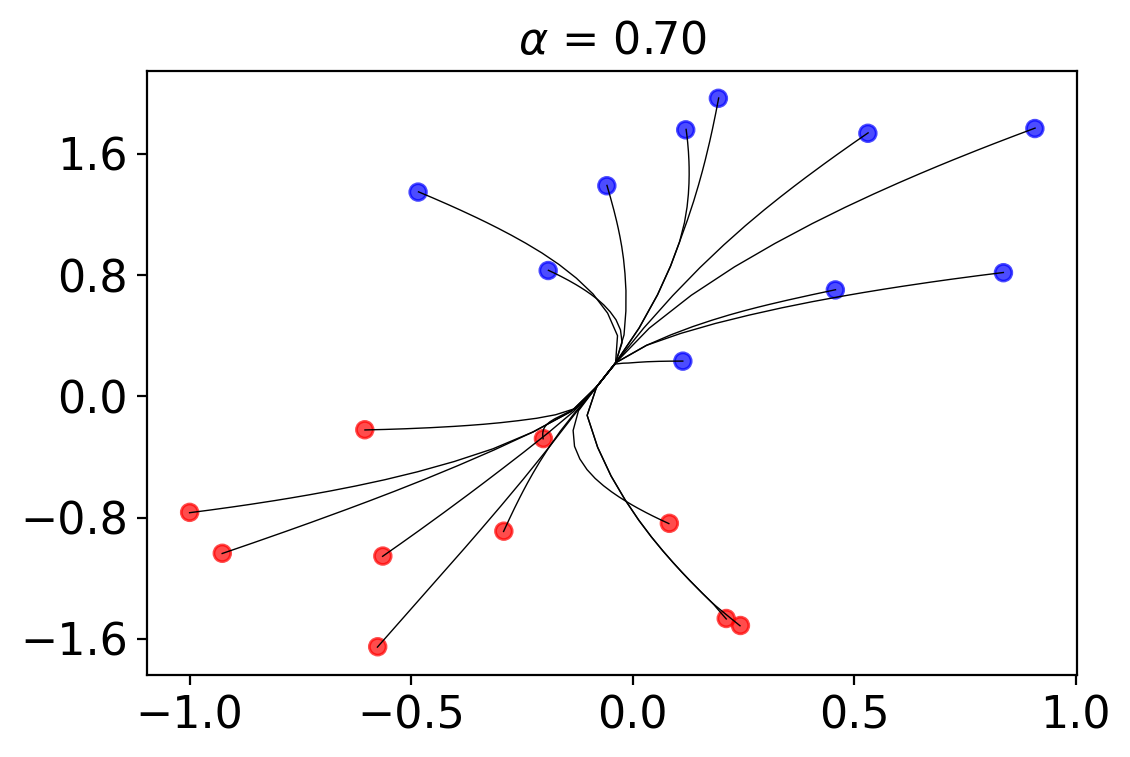} & \includegraphics[width=\linewidth]{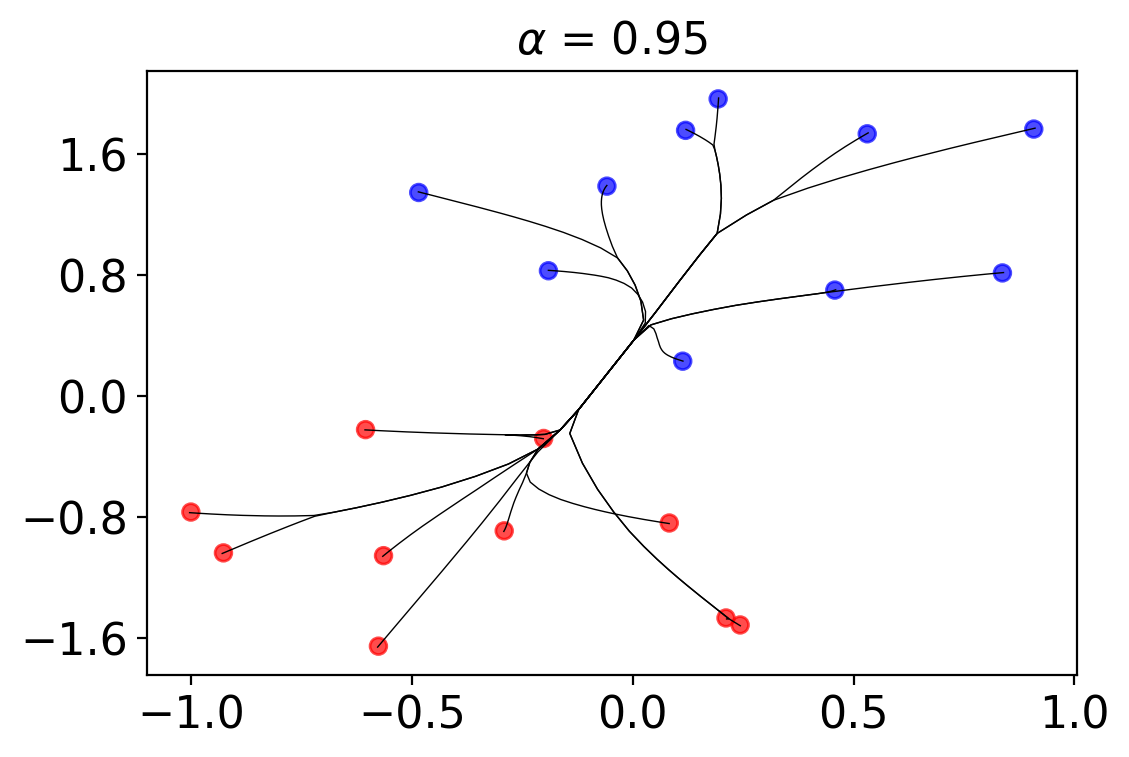} & \includegraphics[width=\linewidth]{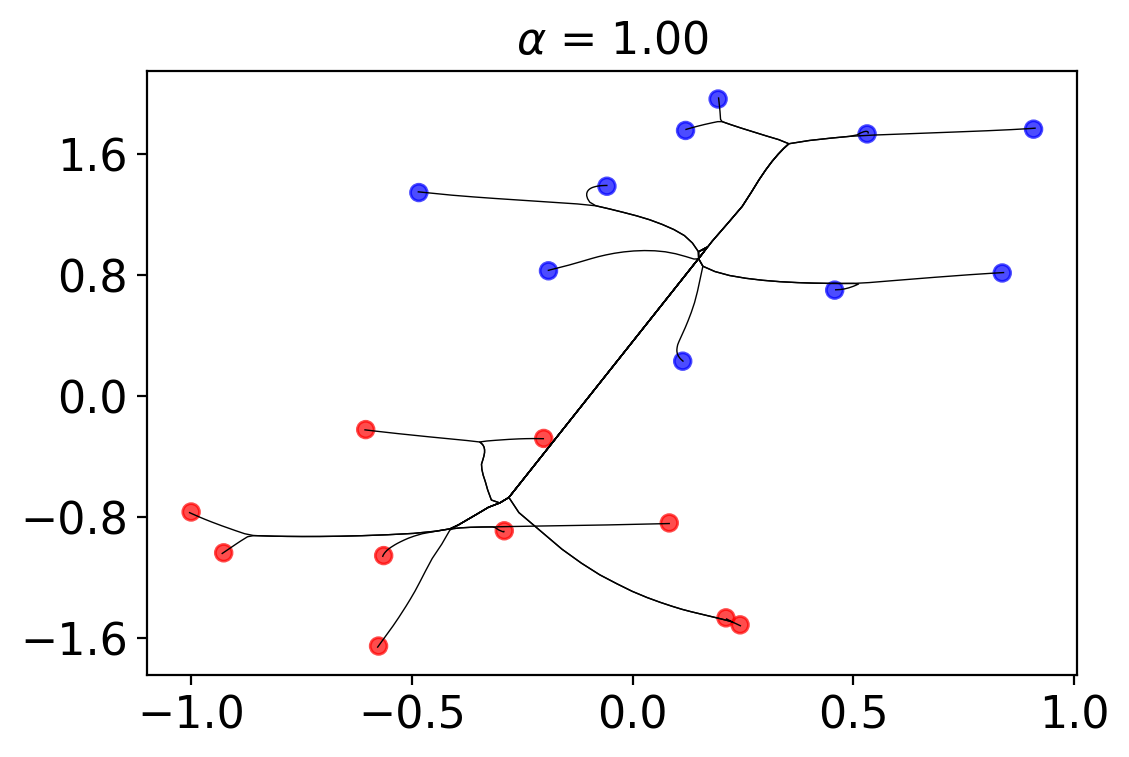} 
        \\

        \\
    \end{tabular}
    \caption{Clustering paths using different weight choices computed on data from two clusters. Top row (uniform weights $w_{ij} = 1$): The graphs show that as the distance between cluster centers increases (left to right), the range of $\gamma$ leading to perfect recovery increases. In the leftmost panel, there is no such $\gamma$ as $\gamma_L > \gamma_U$. Bottom row (convex combination of uniform and Gaussian kernel weights  $w_{ij} = (1 - \alpha) + \alpha w^G_{ij}$): The distance between clusters is fixed. The graphs show that as $\alpha$ increases (left to right), the range of $\gamma$ leading to  perfect recovery increases.}\label{fig:gamma_upper_lower_bound}
\end{figure}

The key insight of \citet{zhu2014convex}, \citet{Panahi2017}, and \citet{jiang2020recovery} is that perfect recovery with uniform weights requires enough separation between clusters to encourage local fusion to occur before global fusion as $\gamma$ increases. Following \citet{Panahi2017}, when $\gamma_L \leq \gamma_U$, there exists a range of $\gamma$ values, $[\gamma_L, \gamma_U],$ where local fusion occurs, but not global fusion. Loosely, $\gamma \geq \gamma_L$ guarantees local fusion {\em within} all $\mathcal{P}_k$, while $\gamma \leq \gamma_U$ guarantees no global fusion {\em across} any two $\mathcal{P}_k$ and $\mathcal{P}_{k'}$. 
\Fig{gamma_upper_lower_bound}, which shows two clusters with ten points each, illustrates this phenomenon. In the first row, the distance between the two clusters in each panel increases from left to right. As the distances increase, the corresponding relationship between the lower and upper bounds transitions from $\gamma_L > \gamma_U$ to $\gamma_L < \gamma_U$. All centroids are shrinking towards the grand mean---global fusion---without any local fusions when $\gamma_L > \gamma_U$, while perfect recovery occurs when $\gamma_L < \gamma_U$.

As we discuss throughout this review, convex clustering is often improved through adaptive weighting schemes. Introducing well-chosen adaptive weights can promote local fusion even when $\gamma_L \gg \gamma_U$. \citet{Sun2021} verify this theoretically by generalizing the results from \citet{Panahi2017} for weighted convex clustering. 
Following \citet{Sun2021}, define


\begin{equation}\label{eq: upper lower bound of gamma, sun2021}
    \begin{aligned}
        \gamma^w_L \amp = \amp & {\max_{k\in [K]}}\  \left\{ \frac{D(\mathcal{P}_k)}{n_k w_{ij} - \mu_{ij}^{(k)}} \right\} \quad \text{ and}\\
        \gamma^w_U \amp = \amp & \underset{1\leqslant k < l\leqslant K}{\min} \left\{ \frac{d(\overline{x}_k,\overline{x}_l)}{n_k^{-1} \sum_{1 \leqslant m \leqslant K, \ m \neq k} w^{(k, m)} + n_l^{-1} \sum_{1 \leqslant m \leqslant K, \ m \neq l} w^{(l, m)} } \right\},
    \end{aligned}
\end{equation}
where 
\begin{equation}\label{eq: gamma bounds aux}
\begin{aligned}
    &w^{(k,m)} \amp = \amp \sum_{i \in \mathcal{I}_k}\sum_{j \in \mathcal{I}_l} w_{ij}, \quad \mu_{ij}^{(k)} \amp = \amp \sum_{l = 1,~ l \neq k}^{K} \left| \sum_{p \in \mathcal{I}_l} w_{ip} - \sum_{p \in \mathcal{I}_l} w_{jp}\right|.
\end{aligned}
\end{equation}

Assuming that $w_{ij} > 0$ for all points $x_i$ and $x_j$ from the same cluster $\mathcal{P}_k,$
\citet{Sun2021} prove that perfect recovery occurs when $\gamma_L^w \leq \gamma \leq \gamma_U^w$. Their result states that perfect recovery is possible if $\gamma_L^w < \gamma_U^w$ with appropriately chosen data-driven weights, even when $\gamma_L > \gamma_U$ using uniform weights. To see a concrete example of this, let us revisit the example shown in the left panel of the top row of Figure~\ref{fig:gamma_upper_lower_bound}. Recall that under uniform weights, $\gamma_L > \gamma_U$. Let us now visualize the effect of smoothly transforming the weights from uniform to Gaussian kernel weights, i.e., for $\alpha \in [0, 1]$
\begin{equation*}\label{eq:convex comb weights}
    w_{ij} \amp = \amp ( 1 -\alpha) + \alpha \,w_{ij}^{G},\quad \quad\text{for all $(i,j) \in \mathcal E$},
\end{equation*}
where $w_{ij}^{G}$ denotes the Gaussian kernel weight. The bottom row of Figure \ref{fig:gamma_upper_lower_bound} displays the solution path for $\alpha \in \{0.70, 0.95, 1.00\}$. We see that centroids of points of the same color fuse before centroids of points of different colors for $\alpha \geq 0.70$. Indeed, there has to be sufficient inter-cluster weight, compared to the between-cluster weight, so that there is within cluster fusion before between cluster fusion.


Other works since \citet{Sun2021} establish perfect recovery conditions in more general contexts. For example, \citet{yagishita2024pursuit} established perfect recovery conditions for the network lasso criterion, which replaces $\sum_{i=1}^n \|x_i - u_i \|_2^2$ with $\sum_{i=1}^n f_i(u_i)$ where $f_i$ are smooth and strongly convex functions. Consequently, these results provide recovery conditions for ``clustered" regression when $f_i(u_i) = \left(x_i - v_i\Tra u_i\right)^2 + (\mu/n) \|u_i\|_2^2$ where $\mu > 0$ is an additional tuning parameter, and $v_i$ are predictors corresponding to $x_i$.

Recently, \citet{dunlap2024sum} establish an insightful negative result under a \textit{two nearby-balls} model. The two nearby-balls model assumes points are drawn uniformly from the union of two disjoint balls of radius one, centered at $r e_j$ and $-r e_j$, respectively, where $r \geq 1$ and $e_j$ is the $j$th standard basis vector in $\Real^p$. They prove that there exists a $\gamma_c$ such that if $\gamma > \gamma_c$, the solution consists of exactly one cluster, whereas if $\gamma < \gamma_c$, the solution consists of at least three clusters as long as $r \in [1, \rho_p)$ where $\rho_p \in [1, \sqrt{2})$ is increasing in $p$. While the result is quite intuitive---it essentially verifies that convex clustering cannot separate points from disjoint balls whose surfaces nearly overlap---the proof requires establishing necessary and sufficient conditions for perfect recovery. To complement their negative result, \citet{dunlap2024sum} show that when $r > 2^{1- 1/p}$, these nearby balls can be separated. They conjecture that the limiting threshold for recovery in the two nearby-balls model is $r = \sqrt{2}$ as $p \to \infty.$


\subsection{Centroid Estimation}
Cluster recovery and centroid estimation are related, though fundamentally distinct.
\citet{Tan2015} provide a general error bound for the recovery of population centers $\{\mu_i\}_{i=1}^n$, under the assumption that $x_i$ is a realization of $X_i = \mu_i + \epsilon_i$ where $\mu_i \in \Real^p$ for $i \in [n]$, and the $\{\epsilon_i\}_{i=1}^n$ are independent and identically distributed sub-Gaussian random vectors with mean zero and covariance $\sigma^2 I_p$.   With uniform weights, \citet{Tan2015} show that for $\gamma \geq 4\sigma \left[\log\{p \binom{n}{2}\}/(n^3p)\right]$, the minimizer of $E_\gamma$, ${u}(\gamma)$, satisfies
\begin{eqnarray*}
\frac{1}{2np}\sum_{i=1}^n \|{u}_i(\gamma)- \mu_i\|_2^2 & \leq & \frac{3 \gamma}{2}\sum_{i<j} \|\mu_i - \mu_j\|_2 + \sigma^2 \left[\frac{1}{n} + \sqrt{\frac{\log(np)}{n^2 p}}\right]
\end{eqnarray*}
with probability tending to one as $n \to \infty$ and/or $p \to \infty.$ If $\mu_i = \mu_j$ for many pairs $i$ and $j$, then it is natural to define the clusters according to the distinct centroids. In the case of few clusters, the bound improves, as many terms in the summation $\sum_{i<j} \|\mu_i - \mu_j\|_2$ vanish.  More generally, if each $\mu_i - \mu_j = \mathcal{O}(1)$, then $\sum_{i<j} \|\mu_i - \mu_j\|_2 = \mathcal{O}(n^2)$. \citet{WanZhaSun2018} later establish a more general result for sparse convex clustering, where an additional 1-norm penalty is added to each $u_i$ in the criterion $E_\gamma$.

\begin{figure}[htbp]
    \centering

    \begin{tabular}{>{\centering\arraybackslash}m{0.08\textwidth} *{3}{m{0.26\textwidth}}}  
        & \makecell{$n = 43$} & \makecell{$n = 79$} & \makecell{$n = 218$} \\
        
        \multirow{2}{*}{\makecell{$2$-\texttt{NNG} \\ + \texttt{MST}}} &
        \includegraphics[width=\linewidth]{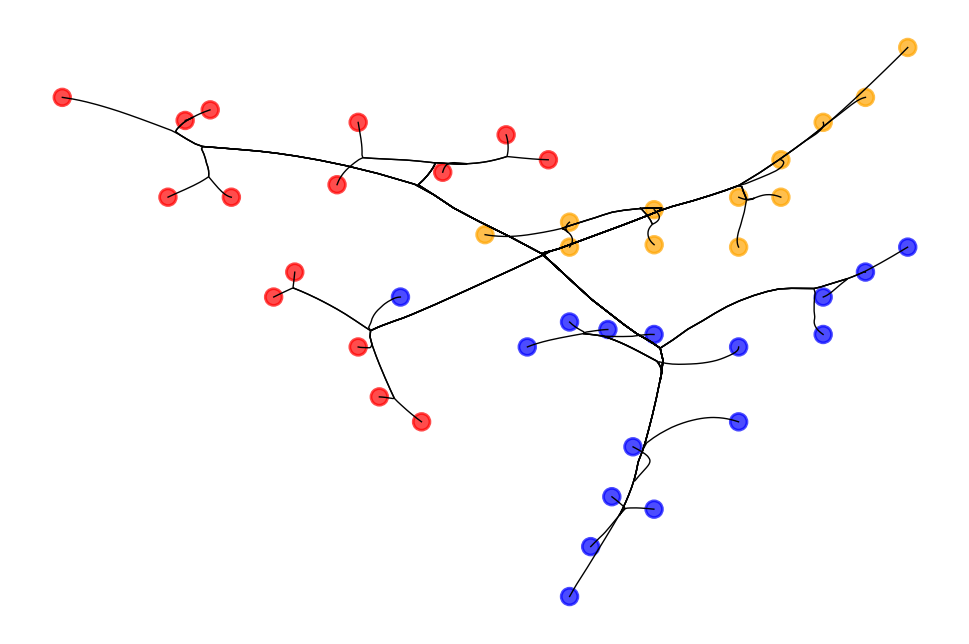} &
        \includegraphics[width=\linewidth]{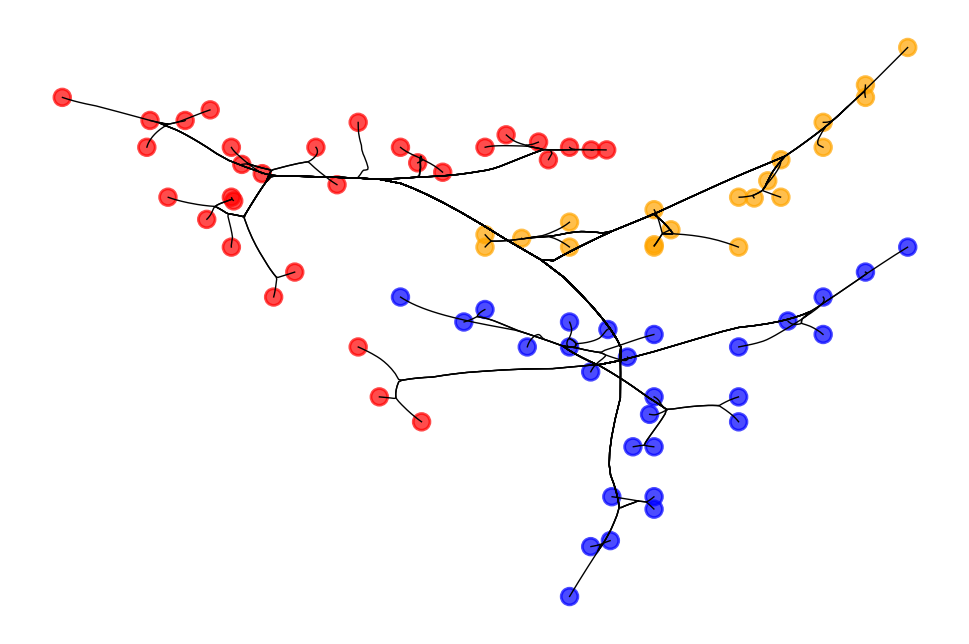} &
        \includegraphics[width=\linewidth]{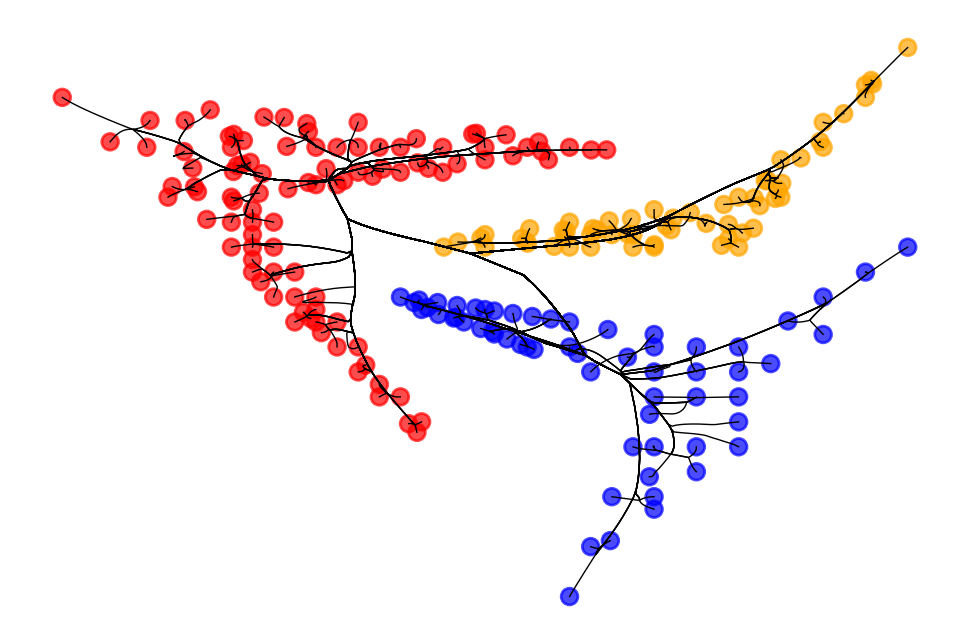} \\
        
        &
        \includegraphics[width=\linewidth]{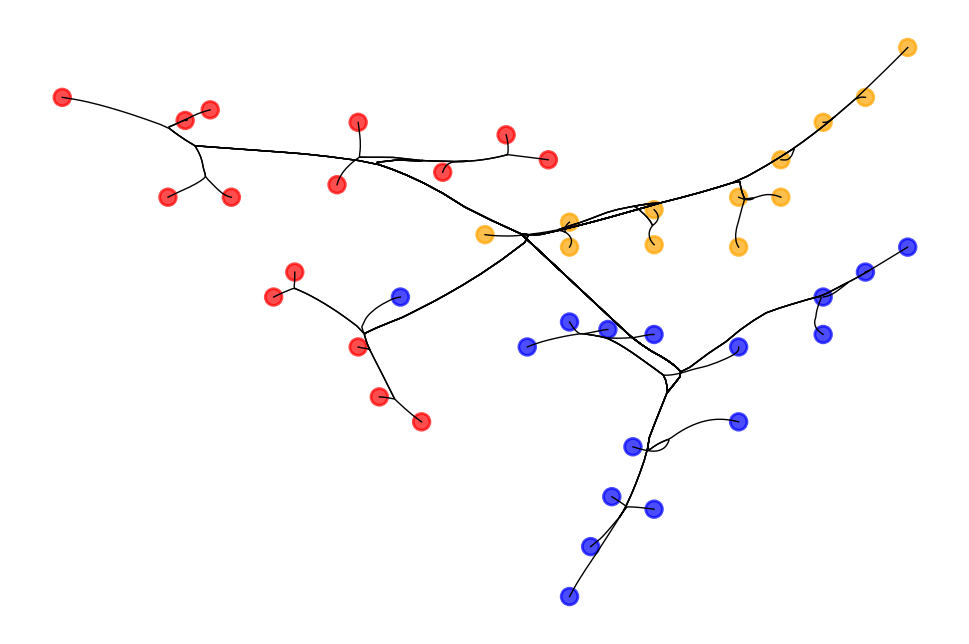} &
        \includegraphics[width=\linewidth]{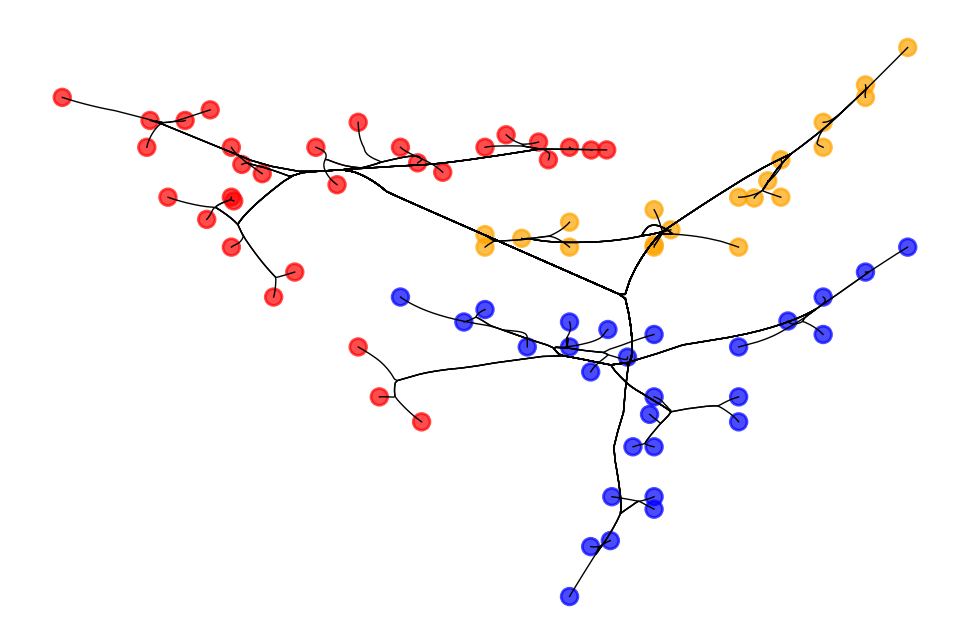} &
        \includegraphics[width=\linewidth]{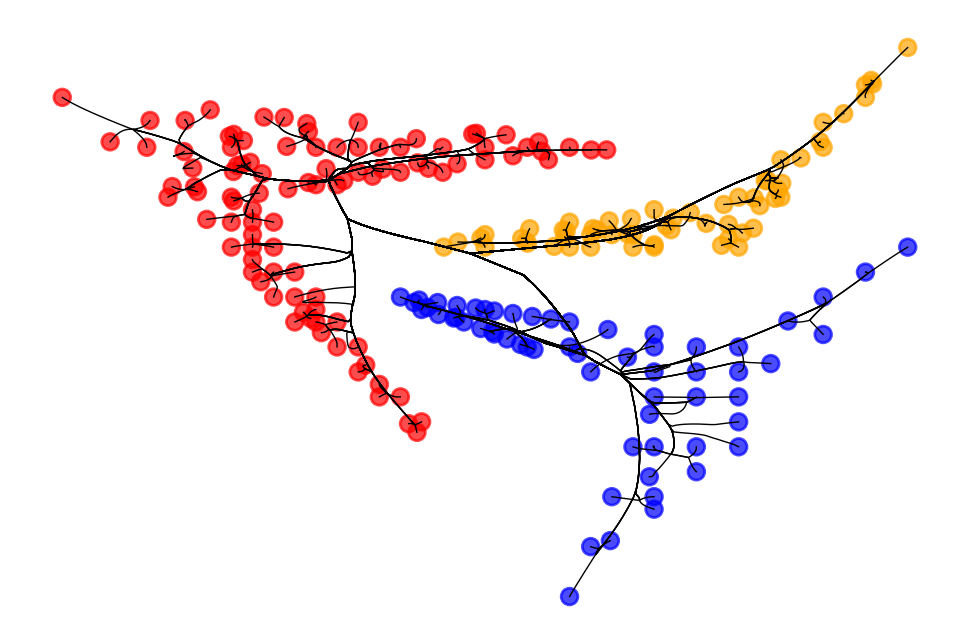} \\
        
        \multirow{2}{*}{\makecell{\texttt{DMSTs} \\ ($M = 3$)}} &
        \includegraphics[width=\linewidth]{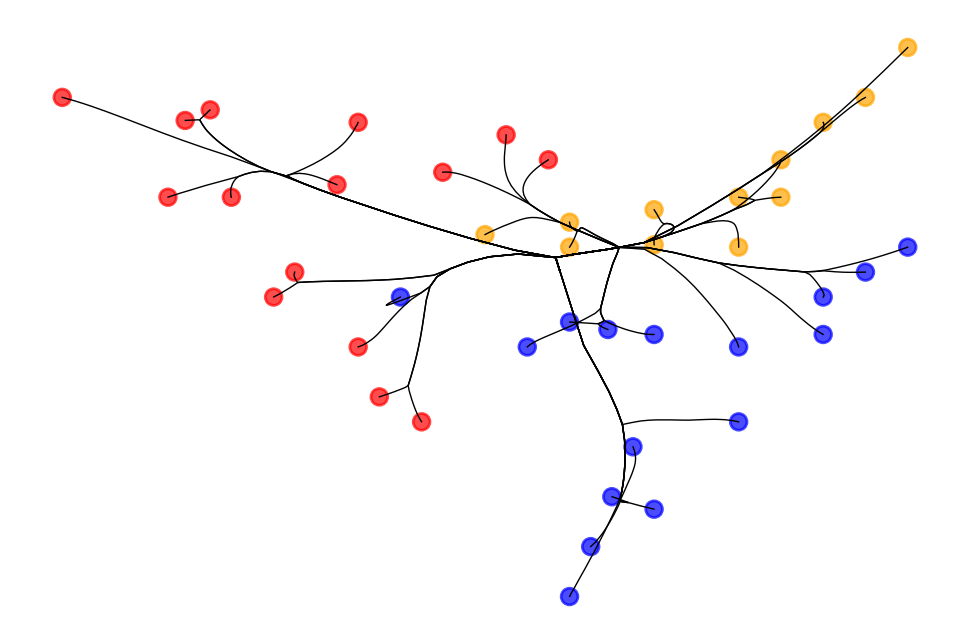} &
        \includegraphics[width=\linewidth]{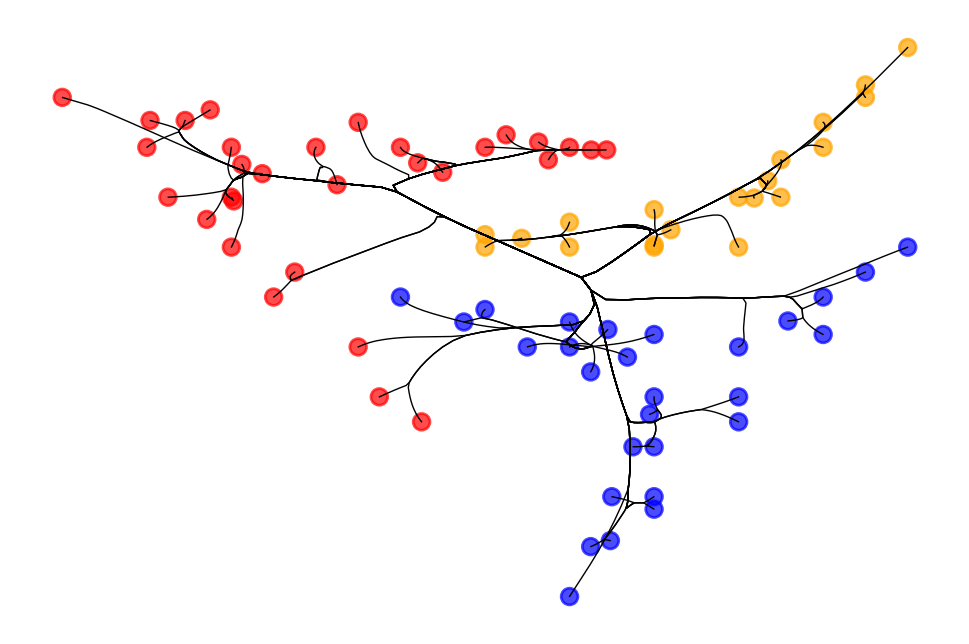} &
        \includegraphics[width=\linewidth]{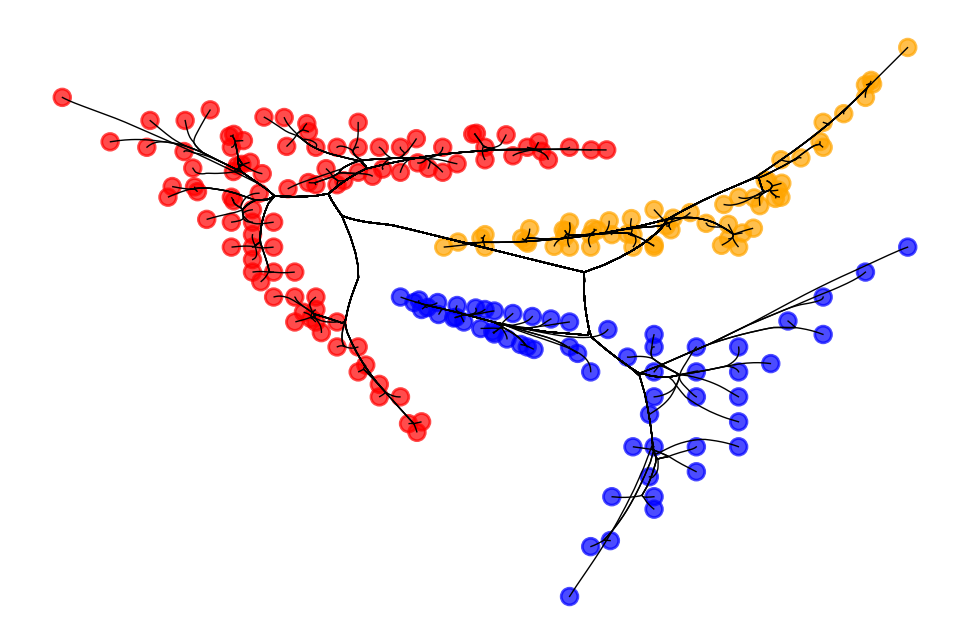} \\
        
        &
        \includegraphics[width=\linewidth]{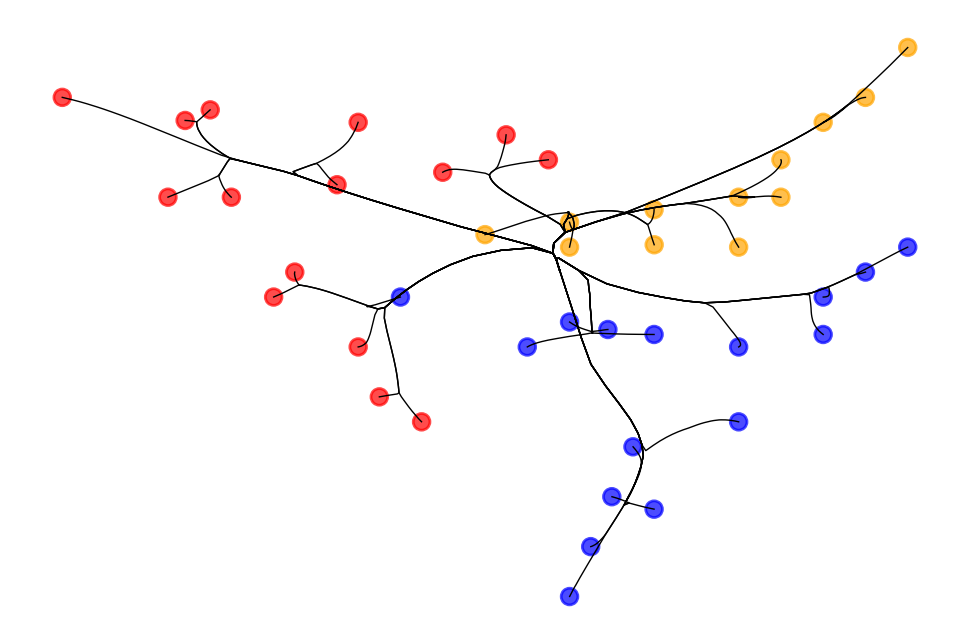} &
        \includegraphics[width=\linewidth]{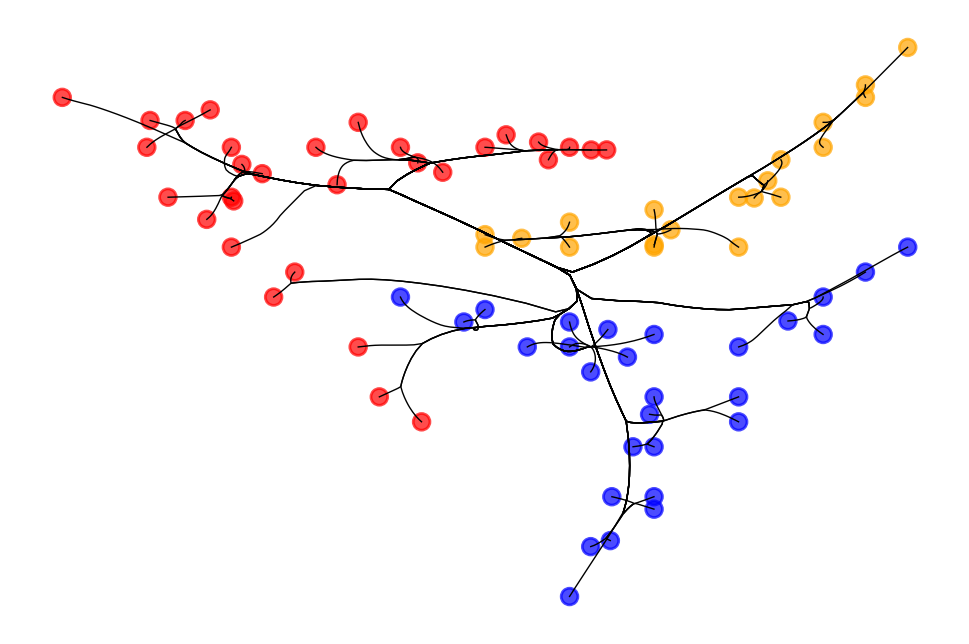} &
        \includegraphics[width=\linewidth]{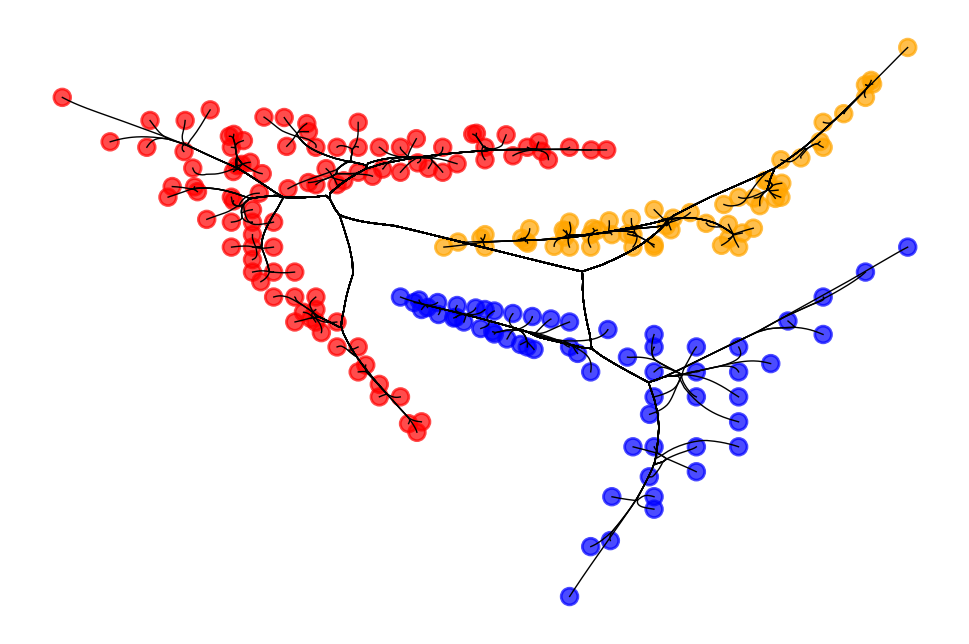} \\
    \end{tabular}

    \caption{Clustering paths on star-shaped clusters with increasing density. Graphs in the top two rows are constructed by $2$-\texttt{NNG} + \texttt{MST}, with inverse Euclidean weights on the first row and Gaussian kernel weights on the second row. Graphs in the bottom two rows are constructed by \texttt{DMST}s $(M = 3)$, with inverse Euclidean weights on the third row and Gaussian kernel distance on the fourth row.}\label{fig:star-shape-paths}
\end{figure}

\citet{Dunlap2022} prove another general result on centroid estimation. They assume that $x_1, \dots, x_n$ are $n$ independent realizations from a continuous distribution with Lipschitz density, whose support is the union of disjoint, closed sets $\mathcal{P}_1, \dots, \mathcal{P}_K$, where each $\mathcal{P}_k \subset \Real^p$ is ``effectively star-shaped'' \citep[Definition 1.1,][]{Dunlap2022}. We refer readers to their paper for the formal definition since it is a bit involved. Suffice it to say, however, that star-shaped sets can be non-convex -- a support geometry that other clustering methods like $k$-means can struggle to navigate as we saw in Figure~\ref{fig:comparisons_starshaped}. 
Define $\psi_k = \mathbb{E}(X \mid X \in \mathcal{P}_k)$ for $k \in [K]$. 
With $w_{ij} = \nu^{p+1}{\rm exp}(-\nu\|x_i - x_j\|_2)$ for a user-specified tuning parameter $\nu > 0$, \citet{Dunlap2022} prove that there exists a $\gamma_c$ such that for all $\gamma \geq \gamma_c$, 
\begin{eqnarray*}
\mathbb{E}\left[  \frac{1}{n}\sum_{i=1}^n \sum_{k=1}^K \mathbf{1}(x_i \in \mathcal{P}_k)\|{u}_i(\gamma) - \psi_k \|_2^2 \right] & \lesssim & \frac{\nu\log (n)^{1/p'}}{n^{1/\max(p,2)}} + \frac{(1 + \gamma)}{\nu^{1/3}},
\end{eqnarray*}
where $p' = \infty \cdot\mathbf{1}(p = 1) + (4/3)\cdot \mathbf{1}(p = 2) + p \cdot \mathbf{1}(p \geq 3).$ When $p \geq 2$, the optimal $\nu \asymp n^{3/4p}$,  which implies that the upper bound is of order $n^{-1/4p}$, up to logarithmic corrections. 
Remarkably, this bound holds for disjoint $\mathcal{P}_1, \dots, \mathcal{P}_K$ that can be arbitrarily close to one another, provided that they are effectively star-shaped.   \Fig{star-shape-paths} shows convex clustering applied to data sampled from densities with effectively star-shaped supports. We see that convex clustering, using Gaussian kernel and inverse Euclidean distance weights, is able to recover non-convex clusters provided sufficiently dense sampling. A notable aspect of their result is that their estimation criterion depends on the data in both the sum of squares term and the weights. Consequently, their analysis requires a particularly delicate treatment compared to those using non-data dependent weights. 

\citet{sun2025resistant} prove a negative result on centroid estimation. They show that if even a single point $x_i$ is adversarially contaminated, convex clustering on the contaminated data can be arbitrarily worse than convex clustering on the uncontaminated data. This lack of resistance to contamination is a consequence of using the sum of squared errors as a loss function. 
Previous efforts to mitigate the impact of outliers that were shown to work empirically include introducing slack variables \citep{Wang2016robust} and metric learning \citep{Sui2018}. 
\citet{sun2025resistant} prove that replacing squared error loss with the Huber loss produces estimates that are resistant to more than half of the points being contaminated.

\subsection{Other Relevant Results}
Other works have considered special cases of convex clustering. For example, \citet{Radchenko2017} study a version of $E_\gamma(u)$ with uniform weights that replaces the Euclidean norm of the difference $\|u_i - u_j\|_2$ with the $\ell_1$-norm of the difference $\|u_i - u_j\|_1$. Since both terms in their objective are separable across the $p$ components of $x_i$ and $u_i$, they focus on the $p = 1$ case. Assuming data are generated from a continuous distribution over the union of disjoint intervals, \citet{Radchenko2017} prove that asymptotically, the solution path (clustering tree) of the sample criterion converges to the solution path of a population-level criterion (both in the number of splits and their locations). Under additional smoothness assumptions on the data-generating density, they show that the sample split points converge to their population counterparts at a rate of $\mathcal{O}(n^{-1/3}).$

In some high dimensional regression problems, one seeks clustered parameter values \citep{Bondell2008, Witten2014,  Price2018}. Variations of convex clustering have also been applied in these regression settings \citep{Ke2015, She2010}. For example, \citet{She2010} studies the clustered lasso criterion where, 
$R_{\gamma_1, \gamma_2}(\beta) = \frac{1}{2}\sum_{i=1}^n (x_i - v_i\Tra \beta)^2 + \gamma_1\sum_{i < j}|\beta_i - \beta_j| + \gamma_2 \sum_{j}|\beta_j|$. 
\citet{She2010} shows that the minimizer of $R_{\gamma_1, \gamma_2}$ consistently recovers the true clusters of the $\beta_i$ under a restrictive irrepresentable-type condition \citep{zhao2006model, lahiri2021necessary}. To partially alleviate this issue, \citet{She2010} suggests replacing $\sum_{i < j}|\beta_i - \beta_j|$ with $\sum_{i < j}w_{ij}|\beta_i - \beta_j|$ for nonnegative weights $w_{ij}$.  



\section{ALGORITHMS}

Substantial effort has been made on designing algorithms for convex clustering since the early work of \cite{Lin2011}, which used the off-the-shelf solver, CVX \citep{cvx}. \cite{Hocking2011} was the first to propose algorithms specific for solving Problem \Eqn{objective_function}. They propose different algorithms depending on the norm used in the regularization term. The 1-norm version can be solved with independent fused lasso solvers, e.g., \cite{Hoefling2010}. The 2-norm version can be solved with the stochastic gradient method. The infinity-norm version can be solved with the Frank-Wolfe method \citep{Frank1956}.

Despite the simplicity of the objective function of Problem \Eqn{objective_function}, efficiently computing its solution is non-trivial. The computational thorn to solving the convex clustering problem is the composition of the non-smooth regularization term with the pairwise differencing operation. Since the pairwise differencing operation is linear, to deal with this thorn many convex clustering algorithms \citep{Chi2015, CheChiRan2015, Yuan18, Sun2021} solve the following equivalent equality constrained problem
\begin{equation}
\label{eq:optimization_problem}
\begin{split}
& \text{minimize}\; \frac{1}{2}\sum_{i=1}^n \lVert x_i - u_i \rVert_2^2 + \gamma \sum_{(i,j) \in \E} w_{ij}\lVert v_{ij} \rVert_2 \\
& \text{subject to $v_{ij} = u_i - u_j$ for $(i,j) \in \E$.}
\end{split}
\end{equation}
Problem \Eqn{optimization_problem} is not the only equivalent way to reformulate Problem \Eqn{objective_function}, e.g., \cite{Hallac2015} and \cite{Yu2025}. Problem \Eqn{optimization_problem}, however, is the formulation behind three of the most commonly used algorithms. These are the Semismooth Newton based augmented Lagrangian method (\texttt{SSNAL}), alternating direction method of multipliers\footnote{Note that \cite{Hallac2015} and \cite{Yu2025} also propose ADMM algorithms but applied to a different equality constrained problem. 
} (\texttt{ADMM}), and alternating minimization algorithm (\texttt{AMA}). All three methods come with convergence guarantees to the global minimizer of Problem \Eqn{optimization_problem}, i.e., regardless of initialization, convergence to the global minimizer is guaranteed \citep{Chi2015, Yuan18, Sun2021}. To date, \texttt{SSNAL}\footnote{Code is available at \url{https://blog.nus.edu.sg/mattohkc/softwares/convexclustering/}} provides the state-of-art run times and scalability to compute the convex clustering solution at a single $\gamma$. Interested readers can find a detailed discussion on the benefits of reformulating Problem \Eqn{objective_function} as Problem \Eqn{optimization_problem} as well as the design differences between these three approaches for solving Problem \Eqn{optimization_problem} in the Supplemental Materials.

An important practical question is how fast are these algorithms or how do they scale with problem size? \texttt{SSNAL} and \texttt{AMA} have per-iteration complexities of $\mathcal{O}(p\lvert \E \rvert)$ \footnote{\texttt{ADMM} in principle can be implemented to have nearly the same per-iteration complexity. See the Supplemental Materials. 
}. Thus, the runtime of the algorithms depends on the sample size $n$ through the edge weights. In the worst case, there is a positive weight $w_{ij}$ between every pair of observations $i$ and $j$, i.e., $\mathcal{O}(\lvert\E\rvert) = n^2$. A nearly fully connected weight graph would be prohibitive, but fortunately, we not only get better computational time but also clustering results by using sparser graphs as seen in \Sec{weights}. Choosing a set $\E$ that is a tree, which has $\mathcal{O}(\lvert \E \rvert) = n$ edges, a disjoint union of $M$ trees, which has $\mathcal{O}(\lvert \E \rvert) = \mathcal{O}(Mn)$ edges, or a $k$-nearest-neighbor graph, which has $\mathcal{O}(\lvert \E \rvert) = \mathcal{O}(kn)$ edges, typically recovers the best clustering results in practice. Thus, deterministic approaches will have per-iteration costs that scale linearly with the size of the data.

In terms of scalability, we need to account for not only per-iteration cost but also the number of iterations. \citet{ho2022structured} studied the so-called ``filtering-clustering'' criterion (i.e., $f(u) + \gamma \sum_{t=1}^T \|D_t u\| $ for discrete difference operators $D_t$) from a computational perspective. They established that the dual of the filtering clustering criterion satisfies a \textit{global error bound} condition, which can imply the linear convergence of first-order methods. As such, their work explains why first order methods, e.g., \texttt{ADMM} and \texttt{AMA} enjoy fast convergence empirically on convex clustering and related problems. Nonetheless, although all three have the same per-iteration complexity, \texttt{SSNAL} is able to capture second order generalized Hessian information at the cost of the first order methods and thus provides the best scalability to date.

Often, the entire solution path is desired and there are computational gains that can be achieved when computing the solution over a sequence of $\gamma$. The solution path's continuity in $\gamma$ suggests employing warm-starts, or using the solution at one $\gamma$ as the starting point for a problem with a slightly larger $\gamma$. This is a well known strategy for saving computations when solving a sequence of lasso-type problems over a grid of smoothing parameters. In fact, this strategy is the basis for the notable computational gains enjoyed by the Convex Clustering via Algorithmic Regularization Paths (\texttt{CARP}) method proposed by \cite{Weylandt2019} when an entire convex clustering tree is desired. We discuss \texttt{CARP} more below and first review two additional computational savings that can be netted along a solution path.

Suppose that for a given $\gamma$, the solution path partitions $\mathcal{X}$ into $K$ clusters $\mathcal{P}_1, \ldots, \mathcal{P}_K$. Let $\overline{u}_k$ denote the centroid of the $k$th cluster, i.e., $u_i = \overline{u}_k$ for all $x_i \in \mathcal{P}_k$. Then Problem \Eqn{objective_function} can be expressed as
\begin{eqnarray}
\label{eq:objective_compressed}
\underset{\overline u \in \Real^{pK}}{\text{minimize}}\ E^{(w)}_\gamma(\overline{u}) & := & \frac{1}{2}\sum_{k = 1}^K n_k \lVert \overline{u}_k - \overline{x}_k \rVert_2^2 + \gamma \sum_{1 \leq k \leq l \leq K} w^{(k,l)} \lVert \overline{u}_k - \overline{u}_l \rVert_2,
\end{eqnarray}
where $w^{(k,l)}$ is the ``total weight" between $\mathcal{P}_k$ and $\mathcal{P}_l$ defined earlier in Equation \eqref{eq: gamma bounds aux}
Under appropriate $w_{ij}$ choices, e.g., as specified in \cite{ChiSteinerberger2019}, if two centroids coincide under a tuning parameter $\gamma_0$, they will continue to coincide for all $\gamma$ greater than $\gamma_0$. Thus, as $\gamma$ increases, the number of distinct variables in Problem \Eqn{objective_function} decreases, and we can express Problem \Eqn{objective_function} as an equivalent weighted convex clustering Problem \Eqn{objective_compressed} with fewer variables. Thus, as $\gamma$ increases, we solve a sequence of increasingly smaller optimization problems \citep{Hocking2011, Zhou2021, Yi2021}.

Adaptive sieving is a technique that generalizes screening rules, e.g., \cite{Tibshirani2011}. In the context of convex clustering, like compression discussed above, sieving aims to solve smaller optimization problems as $\gamma$ increases. The difference is that compression leverages past centroid fusions for a smaller $\gamma$ while sieving predicts future fusions at a larger $\gamma$. \cite{Yuan2022} showed that adaptive sieving can speed up \texttt{SSNAL} by more than 7 times and \texttt{ADMM} by more than 14 times.

We close our review of algorithms by highlighting alternative strategies for computing convex clustering. Stochastic algorithms have been proposed to solve the convex clustering method. \cite{Panahi2017} proposed a stochastic incremental algorithm. The algorithm selects a pair of indices $(i,j)$ uniformly at random from $\E$ and then applies a proximal point update on the terms in objective function of Problem \Eqn{objective_function} that depend on $u_i$ and $u_j$. \cite{Zhou2021} proposed a stochastic parallel coordinate ascent method to solve the dual. For sparse $\E$, e.g.,  generated by a method in Table~\ref{tab:graph_construction}, the dual to Problem \Eqn{optimization_problem} is nearly separable in $z_{ij}$ so dual coordinate ascent updates can be made in parallel safely most of the time when randomly selected.

Algorithms for convex clustering can be parallelized safely to a large degree even if not to an ``embarrassingly parallel" degree. \cite{WuYuan2025} implemented GPU versions of \texttt{SSNAL}, \texttt{ADMM}, and \texttt{AMA}. \cite{CheChiRan2015} employed the inexact proximal distance algorithm, implemented on GPUs, which alternates between highly parallelizable projection steps and centralized dual variable updates. \cite{Fodor2022} posed a variation on convex clustering with a similar sum-of-norms penalty and proposed an ADMM algorithm amenable to computing on an HPC cluster. 

\cite{pi2021dual} proposed convex clustering from the dual of an equivalent epigraph representation of the regularization term. They propose a general framework for convex clustering with additional sparse regularization to solve large-scale problems and provided an efficient first-order algorithm for solving the dual problem.

As noted earlier, \cite{Weylandt2019} propose \texttt{CARP}, which computes convex clustering via a regularization path approach. This is a significantly different tactic than any of the other optimization methods discussed so far. Given a sequence of $\gamma$'s, they combine warm-starts with a single round of \texttt{ADMM} updates. If the sequence of $\gamma$ values are sufficiently close together, they prove that the resulting algorithm regularization path approximates the solution path up to an arbitrarily small error.

Finally, \citet{wang2025euclidean} establish that the convex clustering criterion can be characterized in terms of a ``Euclidean distance matrix" (EDM) criterion \citep{schoenberg1938metric, dattorro2005convex}. This alternative perspective allows for the application of relatively mature computational and theoretical tools from the EDM literature to be applied to the convex clustering problem.


\section{OTHER PRACTICAL CONSIDERATIONS}

There are a number of issues that can arise when employing convex clustering in practice.  Here, we discuss some of the most common ones and  strategies for addressing them.



\subsection{Missing Data}
\label{sec:missing_data}

The convex clustering problem can be solved with minor modifications when $x_i$ is only partially observed. 
Let $\Omega \subset [n] \times [p]$ denote an index set of observed entries of $X$. Then we seek the minimizer of 
\begin{eqnarray}
\label{eq:objective_function_missing}
E_{\gamma}(u; \Omega) & = & \frac{1}{2}\sum_{(l,k) \in \Omega} (x_{lk}-u_{lk})^2 + \gamma \sum_{(i,j) \in \E}w_{ij} \lVert u_i-u_j \rVert_2,
\end{eqnarray}
where $x_{lk}$ and $u_{lk}$ are the $k$th entries of $x_l$ and $u_l$. The modified objective function in \Eqn{objective_function_missing} can be majorized by a function in the form of the objective function of Problem \Eqn{objective_function} by plugging in a current iterate value $\tilde{u}_{ij}$ for $(i,j) \in \Omega^c$ \citep{CheChiRan2015}. The majorization can then be minimized by any complete data convex clustering algorithm. Alternatively, objective function \Eqn{objective_function_missing} can be minimized by making modest modifications to a complete data convex clustering algorithm, e.g., \texttt{SSNAL}.

\subsection{Selecting $\gamma$}

Standard strategies from the penalized regression literature can be used to select $\gamma$ in a data driven manner. We describe three that are commonly used in practice.
First, we can randomly select a hold-out set of elements in the data matrix and assess the quality of a model $u(\gamma)$, or a bias-corrected version of it, based on how well it predicts the hold-out set \citep{Chi2017}. This strategy has been used for principal component analysis \citep{Wold1978} and matrix completion \citep{Mazumder2010}. This approach requires solving a missing data version of convex clustering, which is addressed above in \Sec{missing_data}.  Second, we can use stability selection for clustering  \citep{Wang2010, Fang2012}. This approach comes with selection consistency guarantees and works well empirically \citep{WanZhaSun2018, Wang2021, Wang2023} but can be computationally intensive. 
Third, we can use an information criterion like the BIC using an unbiased estimate of the degrees of freedom \citep{WanZhaSun2018, Tan2015, Chi2018, Majumder2022}. This approach can also come with selection consistency guarantees \citep{Majumder2022} but can be computationally faster than stability selection.

\subsection{High Dimensions}
\label{sec:highdimension}

\cite{Beyer1999} showed that over a broad class of data distributions, as the ambient dimensional increases, distances from a point to its nearest neighbors become indistinguishable from distances to its farthest neighbors. Thus, naively applying convex clustering, a method in which distance metrics play a central role, to high dimensional data 
may perform poorly. Fortunately, many high dimensional datasets can be approximated reliably by a lower dimensional representation or embedding.

In some cases, high-dimensional data consist of many features that contain little to no information about the clustering structure and should be omitted. \cite{WanZhaSun2018} proposed adding a sparsity inducing penalty to the convex clustering problem to address this scenario. More recently,  \cite{Chakraborty2023} proposed a biconvex clustering method that simultaneously estimates centroids as well as feature weights.

In other cases, where there are more nuanced relationships among most or even all the features, we may turn to nonlinear dimension reduction methods \citep{Belkin2003, CoifmanLafon2006, Donoho2003, Tenenbaum2000, Roweis2000}. High-dimensional data encountered in engineering and science can be approximated reliably by a lower dimensional representation. Indeed, manifold learning has proven to be effective as a nonlinear dimension reduction technique in many scientific domains where very high-dimensional measurements are recorded. This is expected since these high-dimensional data are generated from natural processes that are subject to physical constraints and are consequently intrinsically low-dimensional, e.g., conservation laws in physics represent lower-dimensional manifolds in the higher-dimensional state space of possible solutions.

The latter case suggests embedding high-dimensional data into a low-dimensional space, and then computing a convex clustering solution path using the low-dimensional representation of the data. This strategy is especially natural if one uses diffusion maps, since the diffusion distance between two points in high-dimensions can be approximated by the Euclidean distance in the lower dimensional diffusion coordinate space \citep{CoifmanLafon2006}. Once points are embedded in the diffusion maps space, one can use Gaussian kernel affinities and compute the convex clustering solution path using the Euclidean norm in the regularization term.




\section{EXTENSIONS}

Over the past decade, convex clustering and its variant models have been applied to a wide range of application domains, including but not limited to genetics \citep{CheChiRan2015, Fadason2018}, neuroscience \citep{Yao2019, Wang2023}, image processing \citep{bilen2015weakly, Wang2016robust}, and customer segmentation \citep{chu2021adaptive}. Some recent intriguing uses of convex clustering include generating training samples that are graphs \citep{Navarro2023} and estimating transition probabilities of higher order Markov models \citep{Majumder2022}.

In some cases, the original formulation is sufficient but in other cases, convex clustering has been adapted to cluster data where the Euclidean distance is not an appropriate metric to quantify similarity, e.g., binary data \citep{Choi2019}, boxplots \citep{Choi2019B}, histograms \citep{Park2018}, graph-structured data \citep{Donnat2019, Yao2019}, and time series data \citep{Weylandt2021}. There are two ways to deal with this. Sometimes, there may be a natural way to encode the data and its similarities so that applying the original convex clustering model on the transformed data is sensible as in the case of clustering histograms \citep{Park2018}. If there is no such encoding,  there may be natural substitutes for Euclidean distances as in the case of clustering graph-structured data \citep{Donnat2019}.  \cite{Wang2021} provide a systematic treatment of the latter case. They  introduce a flexible convex clustering framework for multi-view data, where multiple sets of potentially very diverse features are all measured on the same collection of samples. Their framework can jointly accommodate a mix of general convex distance metrics, deviances associated with exponential family distributions, and Bregman divergences. 

Convex clustering has also been extended from one-way clustering to biclustering matrices, i.e., simultaneously clustering the rows and columns of a matrix \citep{Chi2017} as well as its generalization to co-clustering tensors \citep{Chi2018}. Convex co-clustering has been adapted for modeling data arrays with structure, e.g., longitudinal \citep{Weaver2024} and compositional data \citep{Wang2023}, and has also been used as a building block for manifold learning methods for array data \citep{michenechicoifmanicml, Zhang2022}.

Encoding clustering structure via a penalty enables relatively straightforward enhancement of other unsupervised learning methods. For example, \cite{buch2024} combine convex clustering with dimensionality reduction techniques to produce hierarchically clustered principal component analysis (PCA), locally linear embedding (LLE), and canonical correlation analysis (CCA). Conversely, formulating clustering as penalized regression also enables a natural form of supervised clustering. \cite{Wang2023} combine generalized linear models with convex clustering within a unified model to leverage label information from auxiliary variables to produce more interpretable clustering.

\section{DISCUSSION}

In a little over a decade, much progress has been made on the theory and applications of convex clustering. Its growing popularity has been catalyzed by advances in provably convergent and efficient algorithms. The simple idea underlying formulating clustering as a penalized regression problem is malleable and has led to natural generalizations to fusion-based clustering of data on different scales as well as innovative ways to incorporate clustering into existing inferential procedures. Expanding on the latter point, a compelling reason to perform clustering simultaneously with other model fitting steps is to avoid ``double dipping." Clustering is often performed within a pipeline that includes other inferential steps. Using the same data throughout the entire pipeline can lead to identification of spurious structure. 

Convex fusion penalties are not the only way to incorporate simultaneous clustering, but they provide a versatile and modular building block with several advantages. If estimation in a procedure is based on minimizing a convex criterion, adding fusion penalties on model parameters maintains the convexity of the criterion, ensuring that all local minima are global. 
These penalized estimators inherit many of the attractive features of the basic convex clustering model, including stability with respect to data perturbations and hyperparameter choices. These penalized estimators empirically also inherit the tree recovery properties of the basic model. Finally, since computing the original convex clustering problem is a proximal map, this makes it easy to build estimation algorithms for these ``cluster-aware" inferential procedures using state-of-the-art algorithms for convex clustering as the main subroutine.

The basic model also accommodates more sophisticated variants of clustering. Metric learning and biconvex clustering are extensions of the basic model that alternate between i) estimating a distance metric between the data and their centroids conditioned on the current centroid estimates and ii) updating the centroids conditioned on the current estimated distance metric. The fusion penalties can similarly be adaptively modified, e.g., adaptive weights to impart robustness  \citep{Shah2017}.


We end by highlighting interesting open questions. Stable tree recovery is one of the main selling points of convex clustering. Empirically, we can recover a desired tree structure in the many extensions of convex clustering from co-clustering of tensors to hierarchically clustered PCA. Tree recovery results similar to \cite{ChiSteinerberger2019} in these more complicated contexts, however, remain to be determined. Even in the case of the basic model there is a gap in our understanding of the weights that guarantee tree recovery. In practice, weights that decay slower than geometrically fast in pairwise distances work well. For example, weights using inverse distances appear to recover trees as well as Gaussian kernel weights but do not require choosing a bandwidth parameter and also lead to weights that shrink nearby points together more aggressively. On a related note, the sufficient conditions for perfect recovery from \cite{Sun2021} could potentially be sharpened, as we found that perfect recovery can be achieved over a wider range of $\gamma$ in practice. Refining their conditions, or even establishing necessary and sufficient conditions---generalizing those from \citet{dunlap2024sum}---would have practical impact. 
Finally, given the importance of weights to tree and partition recovery in theory and practice, we conjecture that accounting for data adaptive weights could improve finite-sample error bounds under common data-generating models. For example, establishing finite-sample error bounds for centroid estimation under the Gaussian mixture model with data-driven weights is non-trivial (owing to the randomness in the weights), but we expect a tighter bound is possible than the general bound from \citet{Dunlap2022} due to the additional parametric structure.

\section{Supplemental Materials}

\subsection{Additional comparisons for case studies}

\begin{figure}[H] 
	\centering
	\includegraphics[width=\textwidth]{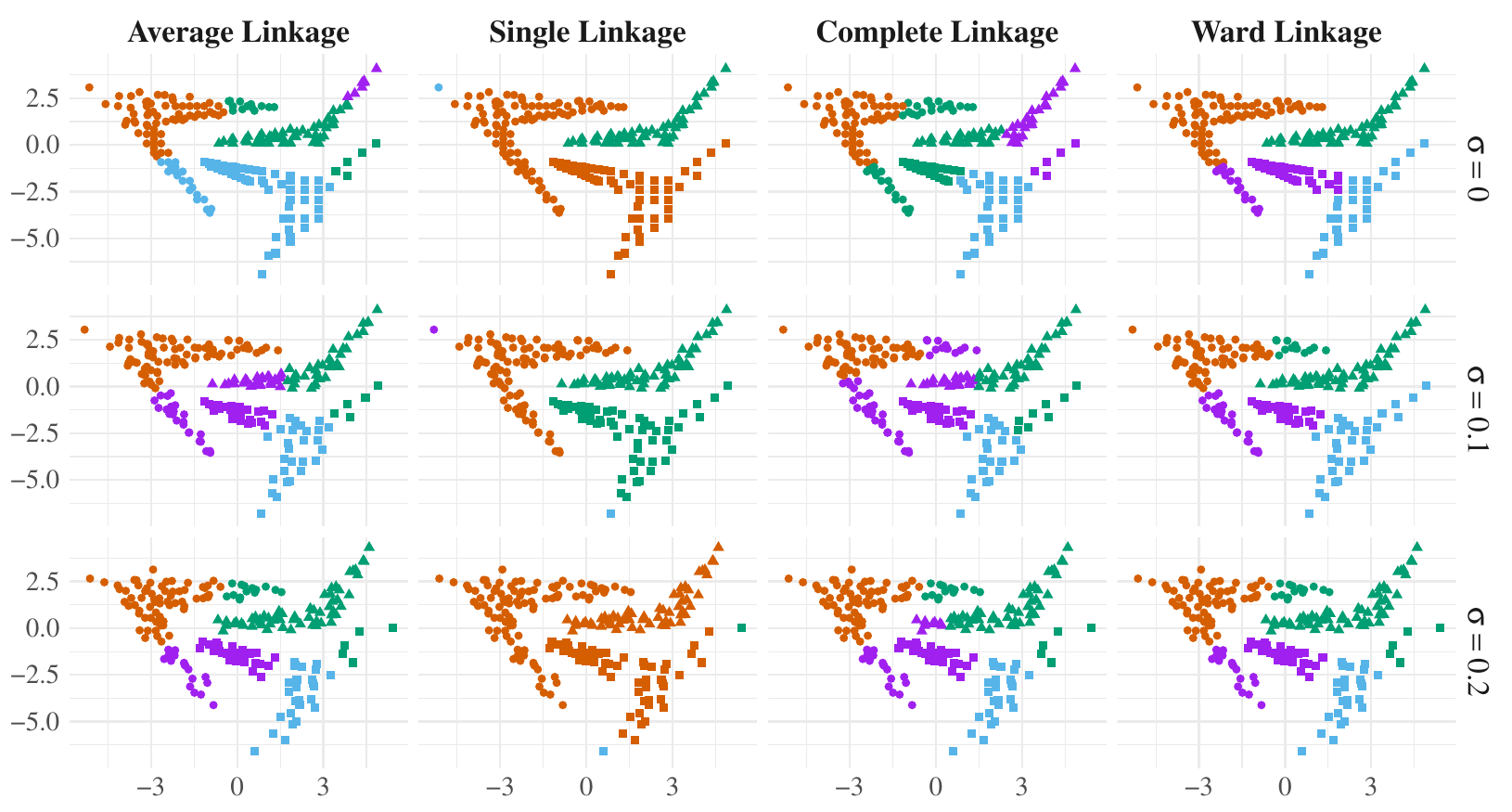}  
	\caption{Experiments on star-shaped dataset highlight differences in hierarchical clustering with average, single, complete, and Ward linkage.}
	\label{fig:comparisons_starshaped_hcall}  
\end{figure}


\subsection{Algorithms}

We expand on the design differences among three of the most often used algorithms for convex clustering: \texttt{SSNAL}, \texttt{ADMM}\footnote{Note that \cite{Hallac2015} and \cite{Yu2025} also propose ADMM algorithms but applied to a different equality constrained problem. In the rest of this section \texttt{ADMM} refers to applying ADMM to Problem \Eqn{optimization_problem}.}, and \texttt{AMA}.

Recall that convex clustering solves the following problem.
\begin{eqnarray}
	\label{eq:objective_function}
	\underset{u \in \Real^{np}}{\text{minimize}}\ \frac{1}{2}\sum_{i=1}^n \lVert x_i-u_i\rVert^2_2 + \gamma \sum_{(i,j) \in \E}w_{ij} \lVert u_i-u_j \rVert_2,
\end{eqnarray}
The three algorithms solve the following equality constrained problem that is equivalent to \Eqn{objective_function}.
\begin{equation}
	\label{eq:optimization_problem}
	\begin{split}
		& \text{minimize}\; \frac{1}{2}\sum_{i=1}^n \lVert x_i - u_i \rVert_2^2 + \gamma \sum_{(i,j) \in \E} w_{ij}\lVert v_{ij} \rVert_2 \\
		& \text{subject to $v_{ij} = u_i - u_j$ for $(i,j) \in \E$.}
	\end{split}
\end{equation}
The purpose of converting the original unconstrained problem to an equality constrained one is to decouple the composition of the pairwise differencing operation and the norm. This is also the motivation for the alternative equivalent reformulations of the original unconstrained problem in \cite{Hallac2015} and \cite{Yu2025}. Decoupling is beneficial because it simplifies the proximal map of the regularizer. 

Recall that the \emph{proximal map} of a function $\psi: \Real^p \rightarrow \Real$ is the operator
\begin{eqnarray}
	\label{eq:proximal_operator}
	\prox_\psi(x) & = & \underset{y \in \Real^p}{\arg\min}\; \left\{\psi(y) + \frac{1}{2}\|y-x\|^2_2\right\}\,.
\end{eqnarray}
The proximal map exists and is unique for all proper, lower-semicontinuous, convex functions $\psi$. Proximal maps are ubiquitous in algorithms for estimating high-dimensional parameters with structured sparsity because norms satisfy these conditions and their proximal maps can often be computed efficiently \citep{ComWaj2005, combettes2011proximal, ParikhBoyd2014, polson2015}. The proximal map for the regularizer in Problem \Eqn{optimization_problem}, i.e., weighted sum of 2-norms, is the group-wise softthresholding operator. This is an analytical expression that can be evaluated in $\mathcal{O}(p\lvert \E \rvert)$ flops. By contrast, the proximal map for the regularizer in Problem \Eqn{objective_function} does not have an analytical form and requires an iterative solver to compute.

Below, we discuss these three deterministic algorithms in some detail to highlight the sources of the differences in computational costs. All three methods  iteratively compute estimates of a saddle point $(U^\star, V^\star, Z^\star)$ of an augmented Lagrangian function $\mathcal{L}_\rho(U, V, Z)$ corresponding to the equality constrained problem \Eqn{optimization_problem}. We want to compute its saddle point because $(U^\star, V^\star)$ is the solution to Problem \Eqn{optimization_problem} and, consequently, $U^\star$ is the unique global minimizer of Problem \Eqn{objective_function}.

Let $U$ denote the matrix in $\Real^{p \times n}$ whose $i$th column is the $i$th centroid $u_i$, $V$ denote the matrix in $\Real^{p \times \lvert \E \rvert}$ whose columns are the centroid difference variables $v_{ij}$, and $Z$ denote the matrix in $\Real^{p \times \lvert \E \rvert}$ of Langrange multiplier variables that correspond to the $\lvert \E \rvert$ equality constraints in Problem \Eqn{optimization_problem}. The Lagrangian for Problem \Eqn{optimization_problem} is
\begin{eqnarray}
	\label{eq:Lagrangian}
	\mathcal{L}(U, V, Z) & = & \frac{1}{2}\lVert X - U \rVert_{\text{F}}^2 + \gamma \sum_{(i,j) \in \E} w_{ij}\lVert v_{ij} \rVert_2
	+ \sum_{(i,j) \in \E} z_{ij}\Tra(u_i - u_j - v_{ij}).
\end{eqnarray}
The augmented Lagrangian is
\begin{eqnarray}
	\label{eq:augmentedLagrangian}
	\mathcal{L}_\rho(U, V, Z) & = & \mathcal{L}(U, V, Z) + \frac{\rho}{2}\sum_{(i,j) \in \E} \lVert u_i - u_j - v_{ij} \rVert_2^2,
\end{eqnarray}
where $\rho$ is a given positive parameter.

Using the variable splitting formulation in Problem \Eqn{optimization_problem} introduces $p\lvert \E\rvert$ parameters, i.e., $v_{ij}$ for $(i,j) \in \E$, to estimate. Thus, we can anticipate any iterative algorithm for solving Problem \Eqn{optimization_problem} has to perform at least $\mathcal{O}(p\lvert \E\rvert)$ arithmetic operations per iteration. Fortunately, we will see that the current state-of-the-art algorithms also perform at most $\mathcal{O}(p\lvert \E \rvert)$ operations per iteration.

\subsubsection{\texttt{SSNAL}}

\begin{algorithm}[th]
	\caption{\texttt{SSNAL}\label{alg:ssnl}}
	\begin{algorithmic}[1]
		\State \textbf{Initialize} $U^0 \in \Real^{p \times n}, V^0, Z^0 \in \Real^{p \times \lvert \E \rvert},$ and $\rho_0 > 0$
		\For{$k = 0, 1, 2, 3, \ldots$}
		\State $(U^{k+1}, V^{k+1}) \gets \underset{U, V}{\arg\min}\; \mathcal{L}_{\rho_{k}}\left(U, V, Z^{k}\right) \amp = \amp \begin{cases}
			U^{k+1} \gets \underset{U}{\arg\min}\; \Phi^{k}(U) \\
			V^{k+1} \gets V^\star(U^{k+1})  \quad\quad\quad\quad\Eqn{ssnl_v_update}\\
		\end{cases}$
		\ForAll{$(i,j) \in \E$}
		\State $z_{ij}^{k+1} \gets z_{ij}^{k} + \rho_{k} \left (u_i^{k+1} - u_j^{k+1} - v_{ij}^{k} \right)$         \EndFor
		\State Update $\rho_{k} > \rho_{k-1}$
		\EndFor
	\end{algorithmic}
\end{algorithm}

\Alg{ssnl} presents pseudocode of \texttt{SSNAL}\footnote{For expositional purposes, \Alg{ssnl} is a simple version of \texttt{SSNAL}. The version presented in \cite{Sun2021} is an inexact ALM algorithm, i.e., steps are allowed to be computed up to some explicit error tolerance.}. 
Note that the $(U, V)$ update in line 3 of \Alg{ssnl} admits a unique solution, since the augmented Lagrangian is strongly convex in $(U, V)$. The optimal $(U, V)$ are computed in two stages via elimination, i.e., for a fixed $U$, we first minimize $\mathcal{L}_{\rho_k}(U, V, Z^k)$ with respect to $V$. The minimizer is 
\begin{eqnarray}
	\label{eq:ssnl_v_update}
	v_{ij}^\star(U) & = & \prox_{\rho_k^{-1}\lVert \cdot \rVert_2}\left(u_i - u_j + \rho_k^{-1}z_{ij}^k\right)
\end{eqnarray}
for all $(i,j) \in \E$. Plugging in the optimal value $V^\star(U)$ into the augmentend Lagrangian gives the following function
\begin{eqnarray*}
	\Phi^k(U) & = & \mathcal{L}_{\rho_k}(U, V^\star(U), Z^k),
\end{eqnarray*}
which is strongly convex and continuously differentiable and therefore minimized at the unique point where $\nabla \Phi^k(U)$ vanishes. Thus, $U^{k+1}$ is the solution to the nonlinear equation
\begin{eqnarray}
	\label{eq:semismooth_root}
	\nabla \Phi^k(U) & = & 0,
\end{eqnarray}
and $V^{k+1}$ is set to $V^\star(U^{k+1})$ according to Equation \Eqn{ssnl_v_update}. 

Although $\Phi^k(U)$ is not twice differentiable, Equation \Eqn{semismooth_root} can be solved by the semismooth Newton method, since the proximal map of a norm is ``strongly semismooth" keeping the core strategy of Newton's method relevant. Computing the Newton direction requires solving a $pn \times pn$ sparse symmetric positive definite linear system which makes it amenable to the conjugate gradient (CG) method. The bottleneck cost in CG is a matrix-vector multiply that requires $\mathcal{O}(p\lvert \E\rvert)$ operations. The linear system is not only sparse but also well-conditioned leading to a superlinear or possibly quadratic convergence rate for CG depending on the data and construction of $\E$. Interested readers can read \cite{Sun2021} for details. 


\subsubsection{\texttt{ADMM}}

\begin{algorithm}[th]
	\caption{\texttt{ADMM}\label{alg:ADMM}}
	\begin{algorithmic}[1]
		\State \textbf{Initialize} $V^0, Z^0 \in \Real^{p \times \lvert \E \rvert}$ and $\rho > 0$
		\For{$k = 0, 1, 2, 3, \ldots$}        
		\State $U^{k+1} \gets \underset{U}{\arg\min}\; \mathcal{L}_{\rho}\left(U, V^k,     Z^k\right)$
		\State $V^{k+1} \gets \underset{V}{\arg\min}\; \mathcal{L}_{\rho}\left(U^{k+1}, V, Z^k\right) \amp = \amp V^{k+1} \gets V^\star(U^{k+1})  \quad\quad\quad\quad\Eqn{ssnl_v_update} $
		\ForAll{$(i,j) \in \E$}
		\State $z_{ij}^{k} \gets z_{ij}^{k-1} + \rho\left( u_i^{k} - u_j^{k} - v_{ij}^{k}\right)$
		\EndFor
		\EndFor
	\end{algorithmic}
\end{algorithm}

\Alg{ADMM} present pseudocode of \texttt{ADMM}. At a high level, there are only two differences between \Alg{ssnl} and \Alg{ADMM}. First, \Alg{ssnl} {\em simultaneously} updates the primal variables $U$ and $V$, while \Alg{ADMM} {\em sequentially} performs block coordinate updates on $U$ and $V$. Second, the parameter $\rho$ is fixed in \Alg{ADMM} while $\rho_k$ increases and must diverge to ensure convergence. The first difference is the primary difference, so we focus on it below.

The most expensive step in \texttt{ADMM} is the $U$-update (line 3 in \Alg{ADMM}) which requires solving the following $n$-by-$n$ linear system. 
\begin{eqnarray}
	\label{eq:admm_U_update}
	UM & = & X + \rho \sum_{(i,j) \in \E} \tilde{v}_{ij}(e_i - e_j)\Tra,
\end{eqnarray}
where
\begin{eqnarray*}
	M & = & I + \rho \sum_{(i,j) \in \E}(e_i - e_j)(e_i - e_j)\Tra \quad\quad\text{and}\quad\quad \tilde{v}_{ij} \amp = \amp v_{ij} + \rho\Inv z_{ij} \quad\text{for all $(i,j) \in \E$}.
\end{eqnarray*}
Due to the multiple right hand sides in Equation \Eqn{admm_U_update}, \cite{Chi2015} propose computing and caching the Cholesky factorization of $M$ to use across repeated $U$-updates in line 3. Using the Cholesky factors leads to $U$-updates that cost $\mathcal{O}(pn^2)$ arithmetic operations, which leads to a substantially more expensive per-iteration cost  than \texttt{AMA}'s as will be explained shortly. This approach, however, ignores the special structure in $M$. The matrix $M$ is a positive symmetric diagonally dominant matrix -- a linear system for which an $\epsilon$-approximate solution can be computed in effort that is nearly linear in the number of nonzero entries in $M$ \citep{SpielmanTeng2004}. Thus, \cite{Chi2015} likely dismissed \texttt{ADMM} as being far inferior to \texttt{AMA} prematurely. With more judicious implementation, the $U$-update in line 3 could be computed approximately in nearly $\mathcal{O}(p\lvert \E \rvert)$ operations. 

The $V$-update for \texttt{ADMM} is identical to the $V$-update for \texttt{SSNAL}. Despite sharing the same $V$-update, however, the resulting $(U^{k+1}, V^{k+1})$ computed by \texttt{SSNAL} and \texttt{ADMM} are different. The \texttt{SSNAL} update likely makes more progress at every iteration since it jointly updates $U$ and $V$ while \texttt{ADMM} updates $U$ and $V$ sequentially.

\subsubsection{\texttt{AMA}}

\begin{algorithm}[th]
	\caption{\texttt{AMA}\label{alg:AMA}}
	\begin{algorithmic}[1]
		\State \textbf{Initialize} $Z^0 \in \Real^{p \times \lvert \E \rvert}$
		\For{$k = 0, 1, 2, 3, \ldots$}        
		\State $U^{k+1} \gets X - Z^k A$
		\For{$(i,j) \in \E$}
		\State $z_{ij}^{k+1} \gets P_{C_{ij}} \left( z_{ij}^{k} + \rho \left[u^{k+1}_i - u^{k+1}_j\right]\right)$
		\EndFor
		\EndFor
	\end{algorithmic}
\end{algorithm}

We can obtain \texttt{AMA} by computing the $U$-update in \texttt{ADMM} (line 3 in \Alg{ADMM}) slightly differently by minimizing the Lagrangian \Eqn{Lagrangian} 
instead of the augmented Lagrangian \Eqn{augmentedLagrangian}
\begin{eqnarray*}
	U^{k+1} & \gets & \underset{U}{\arg\min}\; \mathcal{L}(U, V^k, Z^k).
\end{eqnarray*}
This seemingly trivial change results in a host of simplifications resulting in \Alg{AMA}. The matrix $A \in \Real^{\lvert \E \rvert \times n}$ is the edge incidence matrix corresponding to $\E$, and the product $Z^kA$ in line 3 of \Alg{AMA} can be computed in $\mathcal{O}(p\lvert \E \rvert)$ operations. The mapping $P_{C_{ij}}$ denotes the Euclidean projection onto the set $C_{ij} = \{ z \in \Real^{p} : \lVert z \rVert_2 \leq \gamma w_{ij}\}$ which can be computed in $\mathcal{O}(p)$ operations. These simplifications are well known as \texttt{AMA} has been shown to be performing proximal gradient ascent on the dual problem \citep{Tseng1991}. The proximal gradient method can be accelerated \citep{BeckTeboulle2009, Wright2009, Goldstein2016}.

\section*{DISCLOSURE STATEMENT}
The authors are not aware of any affiliations, memberships, funding, or financial holdings that might be perceived as affecting the objectivity of this review. 

\section*{ACKNOWLEDGEMENTS}

We thank Genevera Allen, Michael Weylandt, Gal Mishne, Chester Holtz, and Xiwei Tang for helpful and interesting discussions. We also thank the anonymous reviewer for helpful feedback in improving this article.
A. J. Molstad's contributions were supported in part by a grant from the National Science Foundation (DMS-2413294).

\bibliographystyle{asa}
\bibliography{references.bib}

\end{document}